\documentclass[aps,prb,twocolumn,floatfix,preprintnumbers,superscriptaddress,amsmath,amssymb,longbibliography]{revtex4-2}
\usepackage{xcolor,bm,lineno,multirow,float,times}
\usepackage[english]{babel}
\usepackage{physics}
\usepackage{rotating}
\usepackage{verbatim}
\usepackage{babel}[english]
\usepackage{graphicx,tabularx}
\usepackage[multiple]{footmisc}
\usepackage[sort&compress]{natbib}
\usepackage[T1]{fontenc}
\usepackage{soul}
\usepackage{dcolumn}
\usepackage{bm}
\usepackage[version=4]{mhchem}
\usepackage{booktabs}
\usepackage{gensymb}
\usepackage{color}
\usepackage[colorlinks=true,urlcolor=blue,linkcolor=blue,citecolor=blue]{hyperref}

\begin{document}

\title{Tunable inter-bilayer magnetic correlations and candidate multipolar physics in the van der Waals oxyhalides DyOCl, DyOBr, and DyOI}

\author{F. C. Brooks}
\email{faith.brooks@gatech.edu}
\affiliation{School of Physics, Georgia Institute of Technology, Atlanta, GA 30332, USA}
\author{X. Bai}
\affiliation{School of Physics, Georgia Institute of Technology, Atlanta, GA 30332, USA}
\affiliation{Department of Physics and Astronomy, Louisiana State University, Baton Rouge, LA 70803, USA}
\author{J. Bacsa}
\affiliation{School of Chemistry and Biochemistry, Georgia Institute of Technology, Atlanta, GA 30332, USA}
\author{V. O. Garlea}
\affiliation{Neutron Scattering Division, Oak Ridge National Laboratory, Oak Ridge, TN 37831, USA}
\author{S. Calder}
\affiliation{Neutron Scattering Division, Oak Ridge National Laboratory, Oak Ridge, TN 37831, USA}
\author{N. Butch}
\affiliation{NIST Center for Neutron Research, USA}
\author{M. B. Stone}
\affiliation{Neutron Scattering Division, Oak Ridge National Laboratory, Oak Ridge, TN 37831, USA}
\author{M. Mourigal}
\email{mourigal@gatech.edu}
\affiliation{School of Physics, Georgia Institute of Technology, Atlanta, GA 30332, USA}

\date{\today}
\begin{abstract}
Rare-earth van der Waals magnets provide a route to combining strong spin-orbit coupling, large magnetic moments, and reduced dimensionality in bulk crystals. We report a comparative study of the dysprosium oxyhalides DyO$X$ ($X=$ Cl, Br, I), which realize square-bilayer networks of Dy$^{3+}$ moments separated by a tunable van der Waals gap. Structural refinements show that increasing the halide ionic radius strongly expands the inter-bilayer spacing while leaving the local bilayer geometry nearly unchanged. Magnetization and heat-capacity measurements reveal two low-temperature anomalies in all three compounds: antiferromagnetic order at $T_{\rm N}\simeq 7$--$10$~K and a broader anomaly near $T_Q \simeq 27$--$30$~K. Single-crystal magnetization on DyOCl and DyOBr establishes a strong hard-$c$-axis anisotropy, consistent with crystal-field analysis of DyOCl, which yields an XY-like ground-state $g$ tensor. Neutron diffraction shows long-range antiferromagnetic order in DyOCl, whereas DyOBr and DyOI exhibit sharp magnetic scattering coexisting with Warren-like diffuse features, consistent with robust in-plane correlations and imperfect inter-bilayer registry. Inelastic neutron scattering on DyOCl identifies crystal-field excitations near $25$--$30$ meV and an additional magnetic mode near 10 meV whose temperature dependence is tied to the high-temperature anomaly. Taken together, these results establish DyO$X$ as a tunable family of quasi-two-dimensional rare-earth magnets and point to candidate multipolar physics associated with low-lying crystal-field states. Direct probes of quadrupolar order, such as resonant x-ray scattering or elastic-constant measurements, will be required to determine the order parameter at $T_Q$.
\end{abstract}

\maketitle
                 
\section{\label{sec:intro} Introduction}

Layered magnetic materials provide a powerful setting in which to tune collective order and excitations by separating strong intra-layer correlations from weaker inter-layer coupling. In van der Waals (VdW) magnets this separation is especially fruitful because bulk crystals remain accessible to thermodynamic, diffraction, and neutron-scattering probes, while weak bonding between layers creates a natural route towards magnetism in reduced dimensions~\cite{lines_magnetism_1969,burch_magnetism_2018}. Most work to date has focused on transition-metal VdW compounds, including the transition-metal halides~\cite{mcguire_crystal_2017,mcguire_cleavable_2020,zhang_antiferromagnetic_2020,zhang_structural_2022}, exfoliable CrI$_3$~\cite{mcguire_coupling_2015,jiang_controlling_2018,noauthor_stacking-dependent_nodate}, and related layered ferromagnets~\cite{chen_magnetic_2013,hu_positive_2022,hu_exchange_2024}. These materials have established VdW magnets as a versatile platform for studying dimensionality, anisotropy, and field-tuned magnetic phases.

Rare-earth VdW magnets offer a complementary route in which large local moments, strong spin-orbit coupling, and crystal-electric-field effects dominate the low-energy physics. Rare-earth oxyhalides and halides, such as $Ln$O$X$ and $LnX_3$ ($Ln=$\! lanthanide, $X=$\! halide), are particularly attractive because their structures and magnetic interactions can be tuned by changing either the magnetic ion or the halide. Historically, these materials have been less explored than transition-metal VdW magnets~\cite{thoma_rare-earth_1965,holsa_thermal_1980}, but recent work revealed a wide range of anisotropic and low-dimensional magnetic behavior~\cite{zhang_anisotropic_2022,wessler_neutron_2022,yang_thermochemistry_2022,zhang_general_2023,wang_structural_2024,pistawala_anisotropic_2024}. Examples include the candidate quantum spin liquid material YbOCl~\cite{zhang_anisotropic_2022,zhang_ground_2024}, as well as the dysprosium oxyhalides DyOCl~\cite{elmaleh_etude_1971,friedt_dysprosium-161_1983,holsa_simulation_1998,holsa_simulation_2000,matas_low_2013,akhtar_comparative_2020,tian_dyocl_2021,chong_synthesis_2022} and DyOBr~\cite{mayer_crystal_1965,holsa_thermal_1980,holsa_interplay_2002,xu_electronic_2020,pan_magnetic_2024}. In particular, DyOCl was recently shown to be an exfoliable rare-earth VdW antiferromagnet with strong magnetic anisotropy and an \(A\)-type ordered state below \(T_N\approx 10\) K~\cite{tian_dyocl_2021}, while DyOBr single crystals exhibit antiferromagnetic order, strong in-plane anisotropy, an additional anomaly near 30 K, and field-induced magnetization plateaus~\cite{pan_magnetic_2024}.

\begin{figure*}[t]
\includegraphics[width=\textwidth]{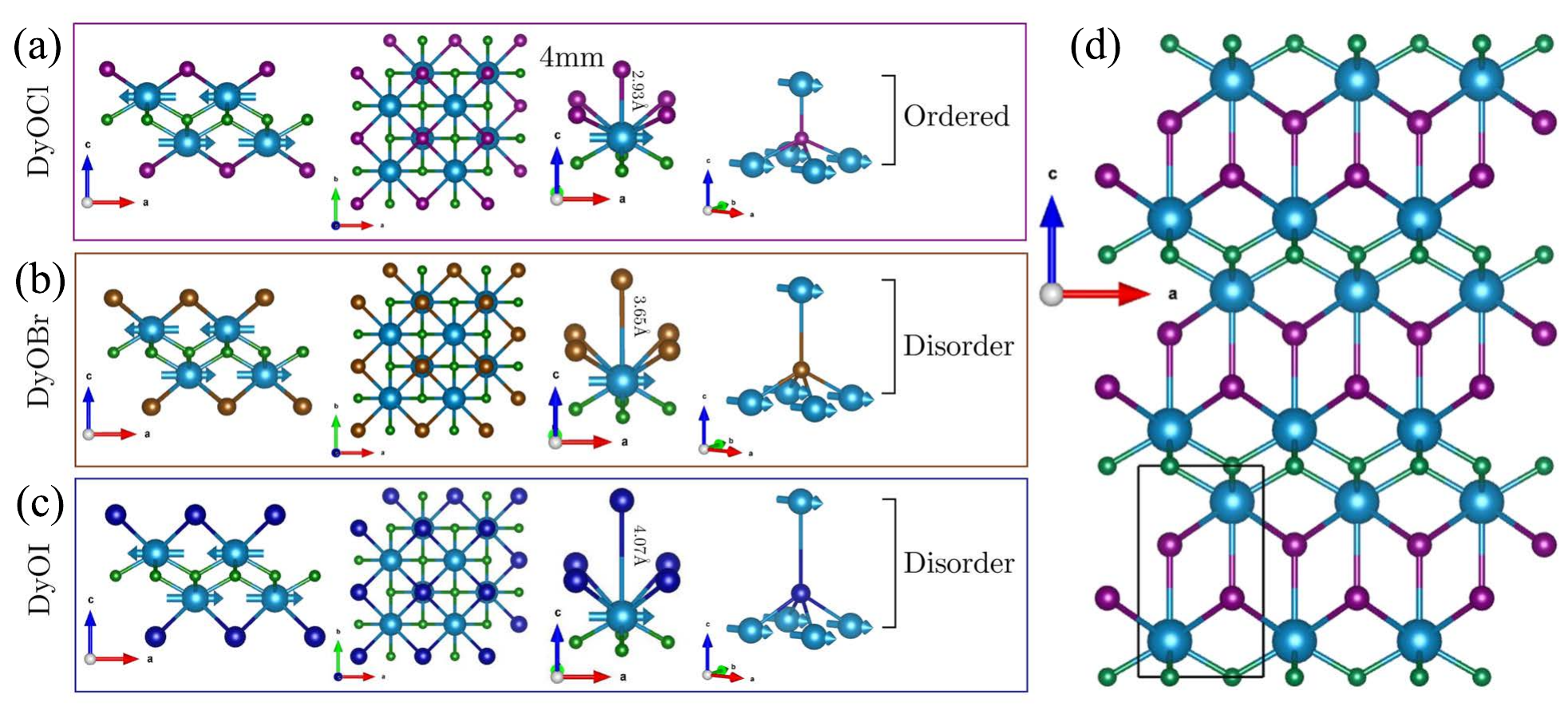}
	\caption{Crystal and magnetic structures of (a) DyOCl, (b) DyOBr and (c) DyOI. The Dy atoms and their magnetic dipole moments are represented by light blue spheres and arrows, respectively. Non-magnetic atoms are represented by green spheres for oxygen, and by purple (Cl), bronze (Br), and sapphire (I) spheres for halides. In each of the panels (a-c), the four subfigures represent, respectively: the bilayer of Dy$^{3+}$ atoms and the associated intra-bilayer magnetic configuration viewed along the $b$-axis; the square lattice arrangement of the Dy$^{3+}$ bilayers viewed along the $c-$axis; the rigid local ligand environment of Dy$^{3+}$ atoms in a bilayer, and the varying inter-bilayer distance; and the relative magnetic stacking between bilayers. (d) Overview of the bilayer stacking with one crystallographic unit cell highlighted. The relative length of the magnetic moments across the three compounds is arbitrary. For DyOBr and DyOI, the illustrated stacking configurations are representative of the correlations inferred from the powder data and are not unique microscopic structures.
	\label{fig:struct}}
	\end{figure*}

The DyO$X$ series ($X=\mathrm{Cl},\mathrm{Br},\mathrm{I})$ is intriguing because it offers chemical control over dimensionality while preserving the same basic magnetic building block. All three compounds crystallize in a tetragonal square-bilayer structure, with Dy$^{3+}$ ions forming bilayers separated by a halide-terminated VdW gap. Increasing the halide ionic radius from Cl to I expands the inter-bilayer spacing substantially, while leaving the intra-bilayer Dy network comparatively rigid. This makes the series a natural platform for separating robust local and intra-bilayer physics from the evolution of inter-bilayer magnetic coupling. In addition, Dy$^{3+}$ carries a large (nominally $J=15/2$) moment and often exhibits strong single-ion anisotropy, making the magnetic behavior highly sensitive to the local crystal-field environment. As shown in Fig.~\ref{fig:struct}, the Dy site has an asymmetric, Janus-like ligand environment, with oxygen and halide ions on opposite sides of the Dy layer, providing a further route for tuning crystal-field anisotropy across the family.

A second motivation for studying DyO$X$ is the possibility of multipolar physics. Several Dy-based materials exhibit quadrupolar or magnetoelastic phenomena~\cite{zaharko_quadrupolar_2004,okuyama_quadrupolar_2005,watanuki_geometrical_2005,yasui_investigation_2009,popova_high-resolution_2017}, often in connection with structural distortions~\cite{zaharko_quadrupolar_2004,okuyama_quadrupolar_2006} or strain coupling~\cite{nakamura_quadrupole-strain_1994,ye_elastocaloric_2022}. More broadly, multipolar degrees of freedom have attracted renewed interest as a route to unconventional ordered and fluctuating states in spin-orbit-coupled magnets~\cite{gao_experimental_2019,massat_field-tuned_2022}. However, quadrupolar order is difficult to establish from thermodynamics alone. In well-established cases such as DyB$_2$C$_2$, direct evidence for antiferroquadrupolar order required resonant x-ray scattering and detailed analysis of the associated superlattice intensities~\cite{hirota_direct_2000,matsumura_observation_2002}. 
For DyO$X$, the higher-temperature anomaly near 30 K motivates a similar multipolar interpretation, but its microscopic order parameter remains to be determined. 

Here we report a comparative study of DyOCl, DyOBr, and DyOI using synthesis, x-ray and neutron diffraction, magnetization, heat capacity, and inelastic neutron scattering. The three compounds share similar bulk thermodynamic behavior, including antiferromagnetic order below $T_{\rm N}\!\approx\!7$--$10$~K and a second anomaly near $T_Q\!\approx\!27$--$30$ K. Single-crystal magnetization on DyOCl and DyOBr establishes strong hard-$c$ axis anisotropy, with additional anisotropy within the basal plane, consistent with crystal-field modeling of DyOCl. Neutron diffraction reveals a sharp contrast between the compounds: DyOCl develops long-range antiferromagnetic order, whereas DyOBr and DyOI show magnetic scattering consistent with robust in-plane correlations but imperfect inter-bilayer registry. Inelastic neutron scattering on DyOCl further identifies crystal-field excitations near 25--30 meV and an additional magnetic mode near 10 meV whose temperature dependence is tied to the high-temperature anomaly. Taken together, these results establish DyO$X$ as a tunable family of rare-earth VdW square-bilayer magnets and point to candidate multipolar physics associated with low-lying crystal-field degrees of freedom.

This paper is organized as follows. Sec.~\ref{sec:methods} details our experimental methods including sample growth, structural and thermomagnetic characterization, and inelastic neutron scattering. Sec.~\ref{sec:results} presents our experimental results and explains the key features of the data. Special emphasis is given to the thermomagnetic, neutron diffraction, and inelastic neutron scattering measurements. Sec.~\ref{sec:discussion} discusses the physical interpretation of the experimental results and the implications of the magnetic diffraction and inelastic neutron-scattering measurements. Sec.~\ref{sec:concl} presents our conclusions and the appendix offers additional miscellaneous information.

\section{\label{sec:methods} Methods}

\subsection{\label{sec:methods:synth} Sample Synthesis} 

Polycrystalline samples of DyOCl, DyOBr and DyOI were synthesized in a manner similar to Ref.~\onlinecite{tian_dyocl_2021}, via solid-state reaction inside capped alumina crucibles placed in a tube furnace under nitrogen gas flow. The reactions proceeded in 1~gram batches according to \ce{Dy_2O_3 (99.99\% Alfa Aesar) + 2NH_4{\it X}  -> 2DyO{\it X} + H_2O + 2NH_3} where $X=\mathrm{Cl}$, $\mathrm{Br}$, or $\mathrm{I}$. Due to the propensity of ammonia to evaporate, excess amounts of NH$_4${\it X} were used in the following ratios: 18\% for DyOCl, 40\% for DyOBr, and 225\% for DyOI. Both DyOCl and DyOBr were heated from room temperature to $T~=~450^{\circ}$C (1 hour), dwelled at $450^{\circ}$C (1 hour), ramped to $700^{\circ}$C (1 hour), and finally dwelled at $700^{\circ}$C (0.5 hour). DyOI was subjected to the same protocol but with a reduced peak temperature of $500^{\circ}$C. Single crystals of DyOCl and DyOBr were synthesized using a \ce{Dy{\it X}_3} flux in capped alumina crucibles under nitrogen gas flow, following the protocol of Ref.~\onlinecite{tian_dyocl_2021}. A 15:1 molar ratio of \ce{Dy{\it X}_3} (99.99\% Alfa Aesar) to \ce{Dy_2O_3} was thoroughly mixed in an alumina crucible heated up to $T~=~1100^{\circ}$C with a 24 h dwell, before cooling to $700^{\circ}$C over 200 h [See Fig.~\ref{fig:app:crystal_pics} for pictures of DyOCl crystals].

\subsection{\label{sec:methods:struc} Structural Characterization} 

Powder x-ray diffraction was performed using a Rigaku SmartLab SE diffractometer with Cu K-$\alpha$ radiation ($\lambda=
1.54056$~\AA). Variable-temperature measurements between $T=300$ and $12$~K were performed using an Oxford Cryosystems PheniX cryostat, with the sample attached to the holder using Molykote grease. Single-crystal x-ray diffraction was performed using a Bruker D8 Venture diffractometer equipped with a HELIOS Cu source and an Oxford Cryostream.

Neutron powder diffraction (NPD) measurements were carried out at the HB-2A beamline at the High Flux Isotope Reactor (Oak Ridge National Laboratory) using $\lambda=2.41$~\AA. Samples were packed in annular aluminum cans with a 1~mm sample thickness to mitigate neutron absorption in natural Dy. The calculated neutron transmission for the annular geometry was $20$~\% at that wavelength. No corrections for absorption were applied. Sample cans were attached to a liquid-helium cryostat, which reached a base temperature of $T=1.5$~K. The diffraction data were rebinned from $2\theta\!=\!5$\degree\, to $127$\degree\, with a spacing of 0.05\degree. Rietveld refinement of the PXRD and NPD data was performed using the FullProf software~\cite{rodriguez-carvajal_recent_1993}.
In all structural refinements we inspected the temperature evolution of peak positions, widths, and potential superlattice reflections to test for symmetry lowering; within our resolution we do not resolve symmetry changes down to the lowest measured temperatures (see details in Sec.~\ref{results:struct}). 

For magnetic structure determination we employed the magnetic space groups approach using the Bilbao Crystallographic Server~\cite{aroyo_crystallography_2011} to enumerate the structures compatible with the nuclear space group $P 4/n m m$ and the propagation vector. When low‑angle lineshapes and broad scattering indicated reduced dimensionality, we analyzed the magnetic diffuse scattering using \texttt{SPINVERT}~\cite{paddison_spinvert_2013} reverse Monte Carlo refinements to obtain real‑space spin–spin correlation profiles (see details in Sec.~\ref{results:dipole}). This workflow allows us to treat magnetic long‑range order and partial/short‑range order on equal footing within the same symmetry framework. The relevant irreducible representations constrain the moments either to the crystallographic $ab$ plane or along the $c$ axis, consistent with the anisotropy result from the crystal-field analysis discussed below.

\subsection{\label{sec:methods:thermo} Thermomagnetic Measurements}

Magnetization measurements were performed on powder samples using the Vibrating Sample Magnetometer option of a Quantum Design Dynacool PPMS. Samples were packed in a plastic cap held by a brass sampler holder; for DyOBr and DyOI, this operation was performed in an argon-filled glovebox to minimize exposure to the atmosphere due to the sample's sensitivity. Temperature-dependent magnetization measurements, $M(T)$, were taken under field cooled (FC) and zero-field cooled (ZFC) conditions in an applied field of $\mu_0H\!\approx\!0.1$~T. Isothermal magnetization measurements were taken up to a magnetic field of $\mu_0H=14$~T. Throughout, the molar susceptibility was calculated as $\chi(T)\!=\!M(T)/H$ and is reported in emu\,mol$^{-1}$\,Oe$^{-1}$. Similar measurements were performed on oriented single-crystal samples of DyOCl and DyOBr. To apply the external magnetic field perpendicular $\perp$ (respectively parallel $\parallel$), to the crystal's $c$ axis, a quartz paddle (respectively a quartz rod) was used. Air-sensitive samples were secured in a Kapton envelope, for which a corresponding background measurement was also collected and subtracted. The mass of the DyOBr sample was not measured directly; instead, the data were scaled so that the saturated magnetization, $M_s\!=\! 10.5~\mu_{\rm B} {\rm Dy}^{-1}$ is consistent with powder measurements (throughout this manuscript, $\mu_{\rm B}=9.274\times10^{-24}$~J/T is the Bohr magneton). This normalization introduces an additional systematic uncertainty in the absolute scale of all DyOBr single-crystal magnetization data.

Heat capacity measurements were performed on pressed, thin powder samples of DyOCl, DyOBr and DyOI on a Quantum Design Dynacool PPMS using the relaxation method. The temperature- and field-dependent specific heat of the samples, $C(T,H)$, was extracted after addenda measurements. Additional measurements were performed on a single crystal of DyOCl oriented using a Laue x-ray backscattering camera. The external magnetic field was applied along the crystallographic $(100)$ and $(110)$ directions using a vertical puck.

\subsection{\label{sec:methods:neutron} Inelastic Neutron Scattering}

Inelastic neutron-scattering measurements were performed on gram-scale polycrystalline DyOCl samples mounted in a flat-slab geometry using an aluminum sachet. Three direct-geometry time-of-flight spectrometers were used. The Disk Chopper Spectrometer of the NIST Center for Neutron Research (NCNR)~\cite{copley_disk_2003} was operated with $E_i\!=\!1.45$~meV which yielded an elastic full-width at half-maximum (FWHM) resolution of $\delta E\!=\!32.1(2)~\mu$eV, suitable for probing putative low-lying magnetic excitations with high resolution. Higher energy excitations were probed using the SEQUOIA~\cite{granroth_sequoia_2010} and HYSPEC~\cite{stone_comparison_2014} spectrometers at the Spallation Neutron Source (Oak Ridge National Laboratory). On SEQUOIA, data were acquired with $E_i\!=\!120$~meV using the high-resolution Fermi chopper spinning at $f\!=\!600$~Hz to achieve an elastic FWHM resolution of $\delta E\!=\!2.90$~meV. HYSPEC was used with $E_i\!=\!3.8$ and 25~meV, and a Fermi chopper frequency of $f\!=\!300$~Hz yielding elastic FWHM resolution of $\delta E\!=\!0.09$ and 1.19~meV, respectively. On DCS, measurements were performed at $T\!=\!11$~K, slightly above the dipolar ordering temperature, using a closed-cycle cryostat; on SEQUOIA and HYSPEC the sample was cooled to below $T = 2$~K using a liquid helium cryostat. No empty-can subtraction was applied to the inelastic neutron-scattering data; the smooth sample-environment contribution was treated as an empirical background.

\section{\label{sec:results} Results}

\subsection{\label{results:struct} Crystal Structure}

\begin{figure}[t]
    \centering
    \includegraphics[width=.45\textwidth]{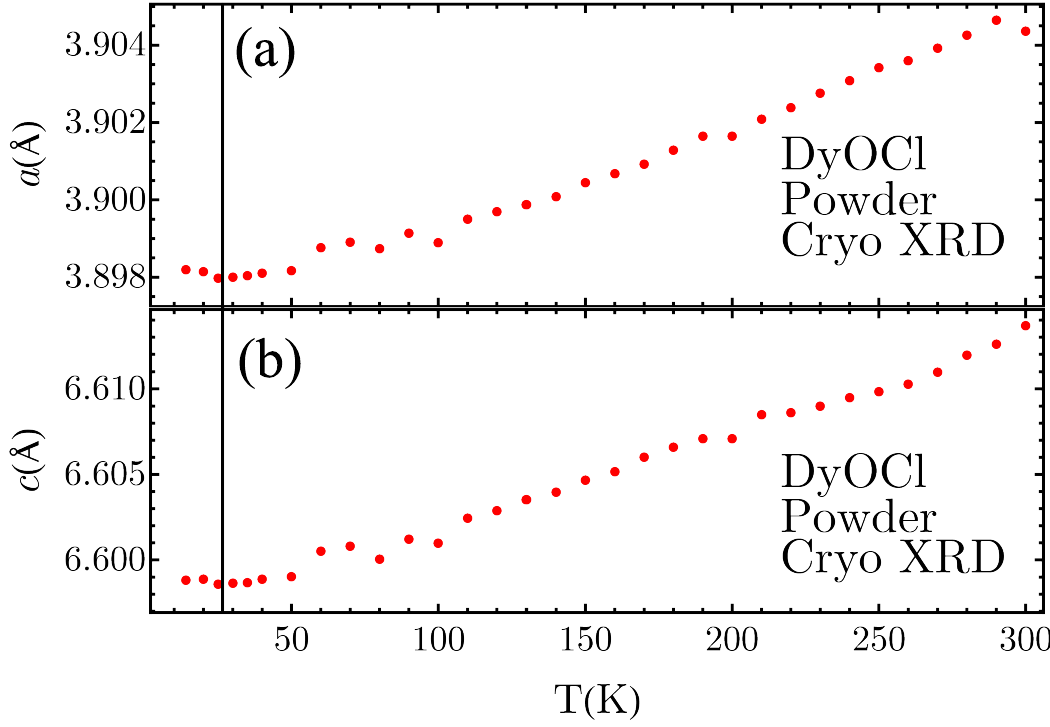}
    \caption{ Temperature dependence of the lattice parameters of DyOCl extracted from powder x-ray diffraction data collected with $\lambda=1.54$~\AA. Values for $a$ and $c$ are found via Rietveld refinement in space group $P 4/n m m$. The vertical line indicates the onset of thermomagnetic transitions in the heat capacity of DyOCl.
    \label{fig:struct:td}}
\end{figure}

All three samples of DyOCl~\cite{elmaleh_etude_1971,friedt_dysprosium-161_1983,holsa_simulation_1998,holsa_simulation_2000,matas_low_2013,akhtar_comparative_2020,chong_synthesis_2022,tian_dyocl_2021}, DyOBr~\cite{mayer_crystal_1965,holsa_thermal_1980,holsa_interplay_2002,xu_electronic_2020,pan_magnetic_2024}, and DyOI studied in this work are expected to crystallize in the tetragonal space-group $P 4/n m m$ (\#129), see Fig.~\ref{fig:struct}, as was most recently confirmed in 2021 for DyOCl~\cite{tian_dyocl_2021} and 2024 for DyOBr~\cite{pan_magnetic_2024}. The phase purity of our powder samples was confirmed by initial PXRD measurements at $T=300$~K, and SCXRD, which produced clean diffraction patterns for all three samples with no unexplained observed peaks [See Fig.~\ref{fig:app:crystal_scxrd}]. Additional temperature-dependent measurements and Rietveld refinements of DyOCl reveal the evolution of the lattice parameters between $T=300$~K and $T=14$~K, as shown in Fig.~\ref{fig:struct:td}. Over this temperature range, the lattice parameters decrease by $\Delta a/a_{\rm 300K} \approx 0.15 \%$ and $\Delta c/c_{\rm 300K} \approx 0.22 \%$. Both decrease smoothly down to $T\approx50$~K and then approach a plateau. No evidence of a structural transition is resolved within the sensitivity of our measurements, including near the $T\approx30$~K thermodynamic anomaly reported below. This contrasts with several Dy-based multipolar materials in which electronic ordering is accompanied by a measurable lattice distortion~\cite{zaharko_quadrupolar_2004,okuyama_quadrupolar_2006,usui_observation_2014}. A detailed analysis of the PXRD data and its temperature dependence is complicated for DyOBr and DyOI, given the extreme air and moisture sensitivity of these samples and the need to use adequate sample holders.
\begin{figure}[t]
    \centering
    \includegraphics[width=.5\textwidth]{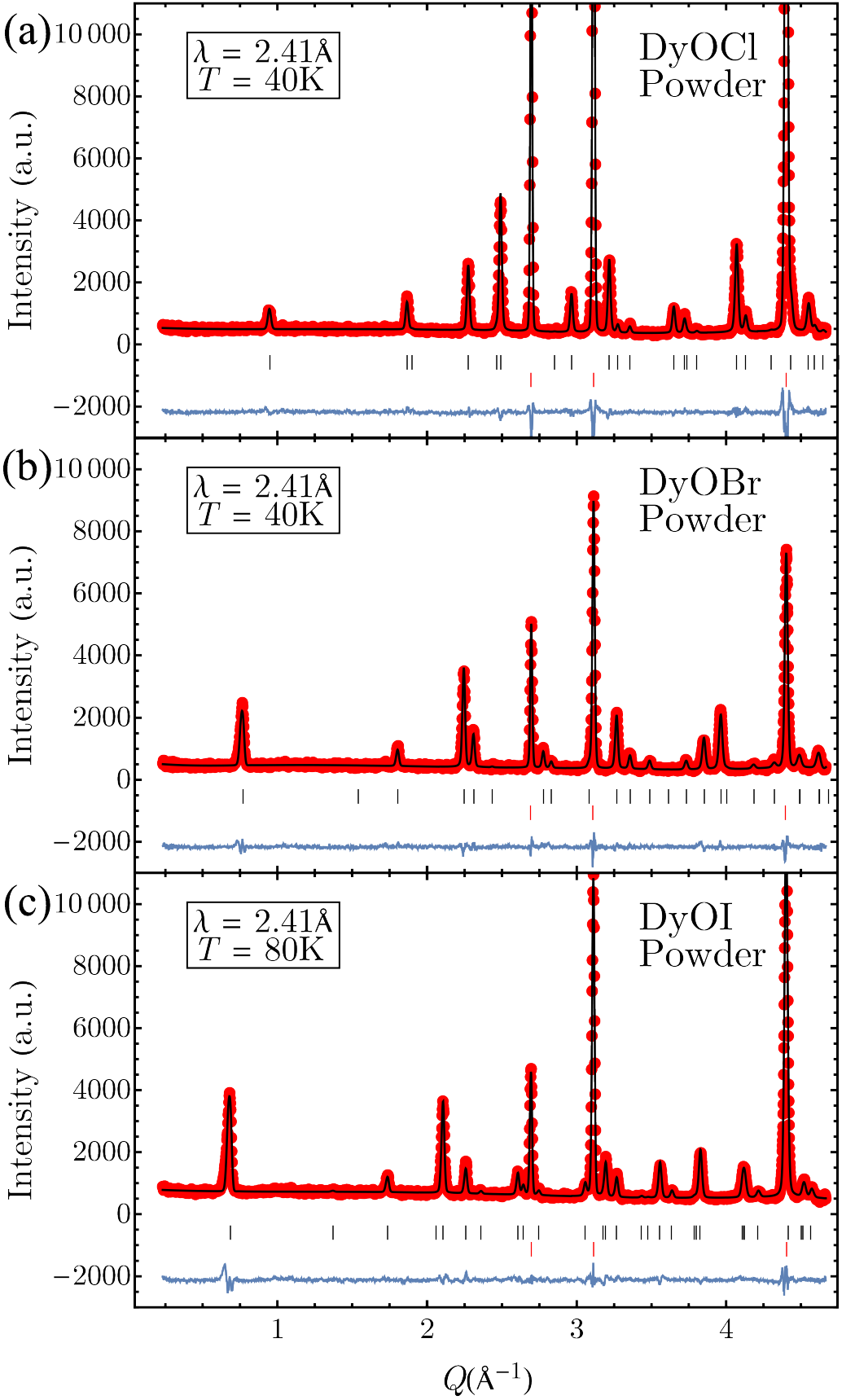}
    \caption{Powder neutron diffraction pattern (red points) for (a) DyOCl, (b) DyOBr, (c) DyOI measured with $\lambda=2.41$~\AA~ in the paramagnetic phase at $T=40$~K or $T=80$~K with corresponding best-fit structural Rietveld refinements (black lines) and residuals (blue lines). Nuclear peak positions are marked as black ticks. Aluminum peaks are marked as red ticks. The residual between the model and the data is marked at the bottom of each panel as a blue curve. We note that low-angle peaks suffer some asymmetry from instrument resolution effects and stacking faults that are not modeled at that stage.
    \label{fig:struct:pararefinement}}
\end{figure}

For a more detailed structural analysis, we turn to neutron powder diffraction results, conducted on all three samples with $\lambda=2.41$~\AA~ at $T=40$~K or $80$~K, above any thermomagnetic transition and cold enough to minimize the Debye-Waller effect. These results and the associated Rietveld refinements are shown in Fig.~\ref{fig:struct:pararefinement}. The model, which only comprises the fractional $z$-position of the lanthanide and the halide as free intra-unit-cell parameters, describes the data well for all three samples, with small residual values. The refined structural information, including lattice parameters and a fit quality metric, is shown in Tab.~\ref{tab:struct:refinement}. The three most intense peaks in each dataset are aluminum from the sample holder; these are modeled independently during fitting. Of particular note is the relative intensity and shift in position of the low-angle peaks across the three compounds, with DyOCl displaying the least intense $(001)$ peak compared to DyOBr and DyOI. This reflects the dramatic 40\% elongation of the $c$-axis from approximatively $6.6$~\AA~ for DyOCl to $9.1$~\AA~ for DyOI while the $a$-axis remains fixed around $3.9$~\AA~ for all three compounds.
\begin{table}[h!]
	\begin{ruledtabular}
		\begin{tabular}{lccccc}
		\textrm{Compound}&
		\textrm{$a=b~(\text{\r{A}})$}&
		\multicolumn{1}{c}{\textrm{$c~(\text{\r{A}})$}}&
		\textrm{$z_{\rm Dy}$}&
		\textrm{$z_{\rm X}$}&
		{R$_{wp}$}\\
		\colrule
		DyOCl & 3.905 & 6.608 & 0.170 & 0.624&11.7\\
		DyOBr & 3.845 & 8.152 & 0.138 & 0.664&15.1\\
		DyOI  & 3.933 & 9.148 & 0.122 & 0.671&12.8\\
		\end{tabular}
	\end{ruledtabular}
	\caption{ Lattice parameters and atomic positions extracted from Rietveld refinement of NPD data for DyOCl, DyOBr, and DyOI at $T=40$~K or $T=80$~K in the space group $P 4/n m m$. Atomic positions are given in fractional coordinates with uncertainties corresponding to the last digit of each number. We note that the refined a-parameter for DyOBr is somewhat smaller than for DyOCl and DyOI which warrants confirmation from single-crystal diffraction.
	\label{tab:struct:refinement}}
\end{table}

With a good structural model derived from NPD, we turn to an in-depth description of the crystal structure of these compounds. Fig.~\ref{fig:struct} shows this structure, with subfigures highlighting relevant structural features. The magnetic Dy$^{3+}$ ions are organized in a face-centered square bilayer geometry, with an oxygen monolayer located between the cations forming the two faces of the bilayer, and halide anions above and below each bilayer, resulting in a van der Walls gap. Within a bilayer, nearest-neighbor Dy$^{3+}$ ions reside on opposite monolayers and are diagonally bridged. The nearest-neighbor interaction ($d_{\rm NN}^X$, for a given halide $X$) is thus inter-monolayer and mediated by two oxygen atoms with a distance around $3.6$~\AA. In contrast, the next-nearest-neighbor distance between Dy$^{3+}$ ions ($d_{\rm NNN}^X$) is intra-monolayer and mediated by two oxygens and two halides with a distance around $3.9$~\AA.

The unique Janus-like ligand environment of the Dy$^{3+}$ ions is asymmetric, with four oxygen anions on one side and four halides on the other. We define the distance between Dy$^{3+}$ cations and their nearest oxygen (respectively halide) anions as $d_{\rm O}^X$ (respectively $d_{\rm H}^X$), while the distance between a Dy$^{3+}$ ion and the nearest halide in the neighboring bilayer (a measure of the van der Walls gap) is defined as $d_{\rm VdW}^X$. These distances are tabulated for all three compounds in Tab.~\ref{tab:struct:distances}. As the halide ionic radius increases from $X$=Cl to I, it is remarkable that bilayers remain very rigid, with changes of $\approx0.4$\% in $d_{\rm O}^X$ and $\approx9$\% in $d_{\rm H}^X$ yielding only minimal changes in $d_{\rm NN}^X$ and $d_{\rm NNN}^X$. On the other hand, the inter-bilayer distance, represented by the van der Walls gap $d_{\rm VdW}^X$, changes by approximately $60\%$. It is also noteworthy that, in the case of DyOCl, the inter-bilayer and intra-bilayer Dy$^{3+}$-halide distances are approximately equal, with $d_{\rm H}^{\rm Cl}/d_{\rm Vdw}^{\rm Cl}\!\approx\!1$, as compared to the other compounds where $d_{\rm H}^{\rm Br}/d_{\rm Vdw}^{\rm Br}\!\approx\!0.75$ and $d_{\rm H}^{\rm I}/d_{\rm Vdw}^{\rm I}\!\approx\!0.67$. This implies that the magnetic ions in DyOCl effectively include an additional ligand in the first coordination shell.
\begin{table}[h!]
	\begin{ruledtabular}
		\begin{tabular}{lcccccc}
			\textrm{Distance $(\text{\r{A}})$}&&
			\textrm{$X=$~Cl}&&
			\textrm{$X=$~Br}&&
			\textrm{$X=$~I}\\
			\colrule
			$d_{\rm O}^X$ && 2.25 && 2.24 && 2.26\\
			$d_{\rm H}^X$ && 3.07 && 3.16 && 3.37\\
			$d_{\rm VdW}^X$ && 3.02 && 4.24 && 5.03\\
			$d_{\rm NN}^X$ && 3.56 && 3.57 && 3.56\\
			$d_{\rm NNN}^X$ && 3.91 && 3.85 && 3.93\\
		\end{tabular}
	\end{ruledtabular}
	\caption{\label{tab:struct:distances}%
	Dysprosium ligand environment distances extracted from refinement of NPD data shown in Fig.~\ref{fig:struct:pararefinement}. Values are uncertain in their last digit.
	}
\end{table}

Overall, our structural analysis identifies DyOCl, DyOBr and DyOI as quasi-two-dimensional magnets with a rigid face-centered square-bilayer motif. From this, we expect the magnetic properties of these samples to feature competing nearest-neighbor (diagonal of the square) and next-nearest-neighbor (side of the square) interactions, as well as pronounced dimensionality and single-ion anisotropy effects as the van der Walls gap increases and the axial crystal-field environment evolves with halide ionic radius. With this understanding, we now turn to thermomagnetic analysis of our samples.

\subsection{\label{results:bulk} Magnetic Properties}

\begin{figure}[t]
    \centering
    \includegraphics[width=.45\textwidth]{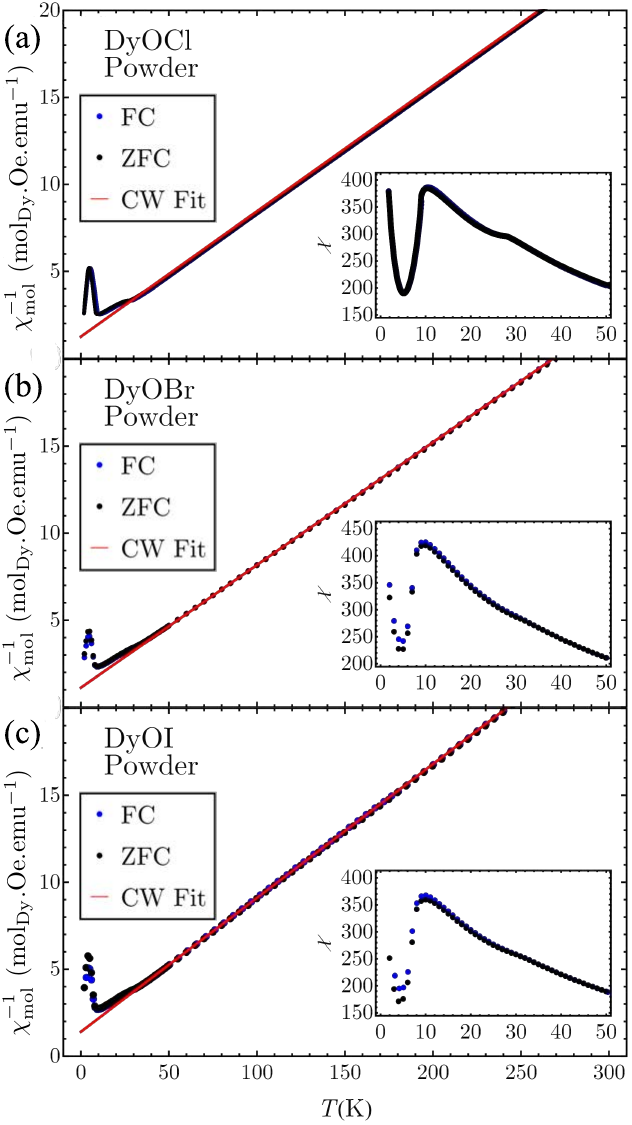}
    \caption{Inverse magnetic susceptibility in a field of $\mu_0 H$=0.1~T as a function of temperature and cooling conditions for each of (a) DyOCl, (b) DyOBr, (c) DyOI. The red curves show fits to the Curie–Weiss law described in the text. The insets show the direct susceptibility over a restricted temperature range near the ordering transitions.}
    \label{fig:mag:chi}
\end{figure}
The temperature-dependent direct and inverse magnetic susceptibilities of our powder samples in a $\mu_0H\!=\!0.1$~T field are shown in Fig.~\ref{fig:mag:chi} and its insets. All three compounds display similar temperature-dependent behavior, with no notable differences between the zero-field cooled (ZFC) and field-cooled (FC) conditions. The direct susceptibility $\chi(T)$ shows the typical increase with lowering temperature, with two notable features at low temperatures in each of our samples. First, a sharp peak around $T\!\approx\!10$~K is followed by an abrupt decrease in $\chi(T)$. We associate this behavior with the antiferromagnetic ordering of all three compounds at the N\'eel temperature $T_{\rm N}$. Second, there is a distinct kink just below $T\!\approx\!30$~K, which is most visible in DyOCl but can also be seen in DyOBr and DyOI. We will discuss the nature of this feature, which is also seen in Ref.~\onlinecite{tian_dyocl_2021}, further below. We denote this temperature $T_Q$ for consistency with the discussion below, while emphasizing that the microscopic order parameter associated with this anomaly is not determined by susceptibility alone. At the lowest temperatures, all compounds show a moderate increase in $\chi(T)$ consistent with a Curie tail. The value of $T_{\rm N}$ for all three samples is determined from the minimum in $|{\rm d} \chi(T) / {\rm d}T|$ and reported in Tab.~\ref{tab:mag:parameters}.

\begin{figure}[t]
    \centering
    \includegraphics[width=.47\textwidth]{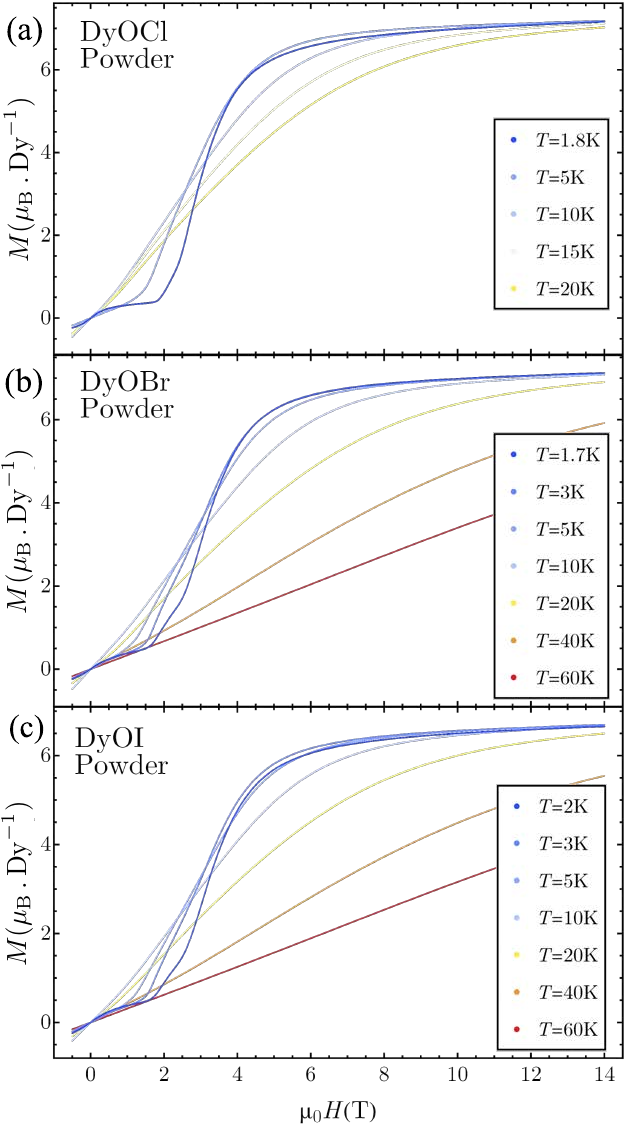}
    \caption{Isothermal magnetization for selected temperatures between $T=1.7$~K and 60~K measured on powder samples of (a) DyOCl, (b) DyOBr and (c) DyOI.}
    \label{fig:mag:isomagpow}
\end{figure}
Above $T\!=\!50$~K, the inverse susceptibility curve appears linear for all three materials. We analyze it using the Curie-Weiss law $\chi (T) = C/(T-\theta_{\rm W}) + \chi_0$ where $C$ and $\theta_{\rm W}$ are the Curie and Weiss constants and $\chi_0$ is a temperature-independent contribution. Fits to the data between $T=50$ and 300~K yield negative Weiss temperatures $\theta_{\rm W}\!=\!-17.2$~K (Cl), $-20.4$~K (Br), and $-16.7$~K (I), indicating overall antiferromagnetic interactions, consistent with the shape of $\chi(T)$ close to $T_{\rm N}$. The corresponding effective moments $\mu_{\rm eff} \approx \sqrt{8C}\mu_{\rm B}$ yield $10.5~\mu_{\rm B}$/Dy , $10.5~\mu_{\rm B}$/Dy, and 9.8~$\mu_{\rm B}$/Dy for DyOCl, Br, and I respectively. These values are close to the free-ion moment expected for Dy$^{3+}$ ($4f^9$, $^6H_{15/2}$, $g_J = 4/3$) from Hund's rules, $\mu^{\rm free}_{\rm eff}\!=\!g_J \sqrt{J(J+1)}\!=\!10.62~\mu_{\rm B}$, with a notable 8\% reduction for DyOI. Overall, the analysis yields ratios $f=|\theta_{\rm W}/T_{\rm N}|$ of 2.0, 2.9, and 2.4 for DyOCl, Br, and I, respectively, implying some degree of magnetic frustration in each of the compounds, see Tab.~\ref{tab:mag:parameters}.
\begin{table}[b]
    \begin{ruledtabular}
        \begin{tabular}{lccccc}
        \textrm{Comp.}&
        $T_{\rm N}$ (K)&
        \textrm{$\mu_{\rm eff} (\mu_B)$}&
        \multicolumn{1}{c}{$\theta_W (K)$}&
        \textrm{$|\theta_W |/ T_{\rm N}$}&
        \textrm{$M_{\rm sat}$ ($\mu_B$/Dy)}\\
        \colrule
        DyOCl& 8.84 & 10.5 & -17.2 & 2.0 & 10.5 \\
        DyOBr& 6.95 & 10.5 & -20.4 & 2.9 & $\ast\ast$ \\
        DyOI & 6.85 & 9.8  & -16.7 & 2.4 &  - \\
        \end{tabular}
    \end{ruledtabular}
\caption{Tabulation of magnetic properties for DyOCl, DyOBr, and DyOI extracted from susceptibility and magnetization measurements. Values are uncertain in their last digit. $\ast\ast$ Due to uncertainty on the sample mass, the DyOBr data were normalized to reach 10.5~$\mu_{\rm B}$ at saturation. \label{tab:mag:parameters}}
\end{table}

We now focus on the isothermal magnetization versus external field for powder samples, $M^{\rm pow}(H)$, shown in Fig.~\ref{fig:mag:isomagpow}. At the lowest temperatures, the magnetization does not fully saturate even at $\mu_0H\!=\!14$~T, but asymptotically plateaus around $M^{\rm pow}_s\!\approx\!7$--$8~\mu_{\rm B}/{\rm Dy}$. This value is well below the expectation for Heisenberg-like Dy$^{3+}$ free moments, $M_s\!=\!g_J J\!=\!10~\mu_{\rm B}$, which is otherwise confirmed by susceptibility measurements. We tentatively assign this reduction to powder averaging of anisotropic moments: only crystallites with favorable orientations contribute efficiently to the high-field moment. At lower fields, all three samples exhibit the same low temperature, highly non-linear behavior. In this region, all three show notably small magnetization before a jump at $\mu_0H\!=\!1.79$~T (DyOCl), $1.57$~T (Br), and $1.61$~T (I), after which the magnetization generally increases rapidly with applied magnetic field. We interpret these signatures as field-induced rearrangements of antiferromagnetically ordered moments. Taken together, the incomplete high-field saturation and the low-field jump-like behavior indicate strong magnetic anisotropy.

\begin{figure}[t]
    \centering
    \includegraphics[width=.5\textwidth]{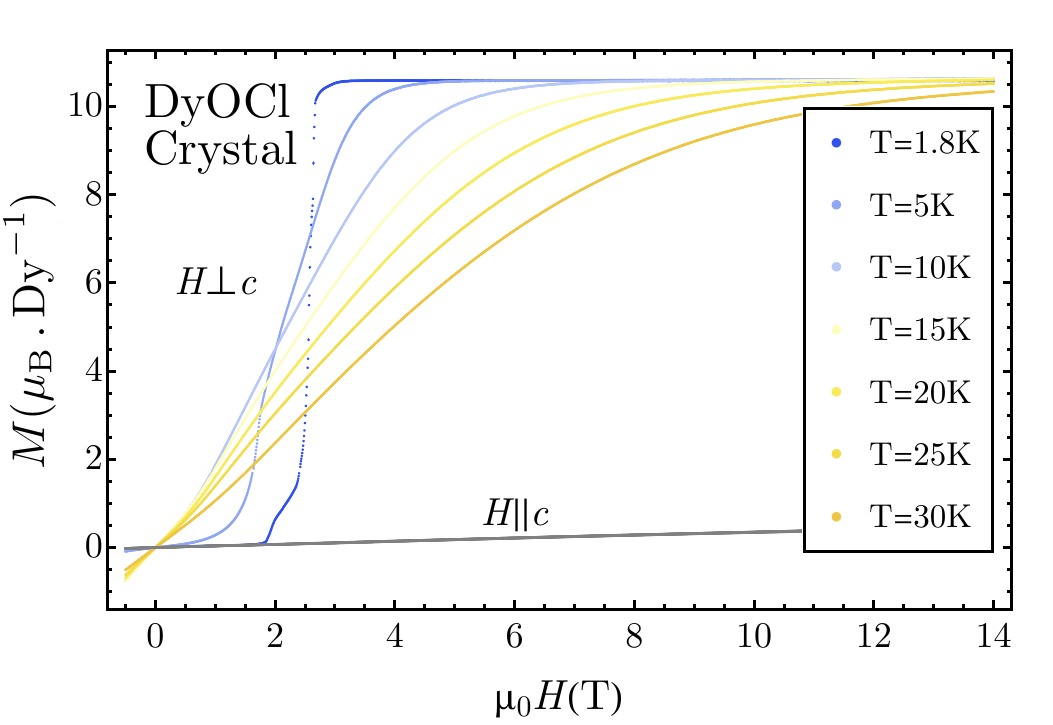}
    \caption{Isothermal magnetization on a single crystal sample of DyOCl measured with the applied magnetic field parallel and perpendicular to the crystallographic $c$-axis. The orientation of the field in the $ab$-plane is not known.}
    \label{fig:mag:isomagcl}
\end{figure}

\begin{figure}[t]
    \centering
    \includegraphics[width=.5\textwidth]{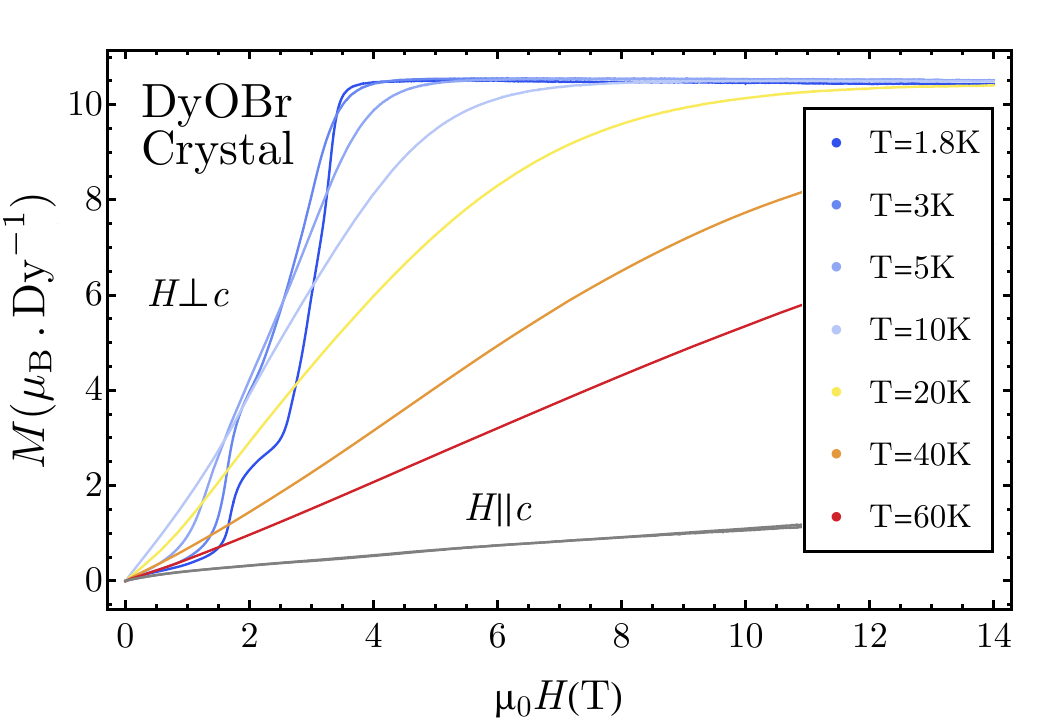}
    \caption{Isothermal magnetization on a single crystal sample of DyOBr measured with the applied magnetic field parallel and perpendicular to the crystallographic $c$-axis. The orientation of the field in the $ab$-plane is not known.}
    \label{fig:mag:isomagbr}
\end{figure}

To resolve the magnetic anisotropy directly, we turn to isothermal magnetization measurements on oriented single crystals of DyOCl [Fig.~\ref{fig:mag:isomagcl}] and DyOBr [Fig.~\ref{fig:mag:isomagbr}]. For magnetic fields applied perpendicular to the crystallographic $c$-axis, $H\perp c$, the magnetization rises much more rapidly than in powder-averaged measurements and approaches the full Dy$^{3+}$ moment, $M\simeq 10.5~\mu_{\rm B}/{\rm Dy}$, by $\mu_0H\simeq 3.5$~T at $T=1.8$~K. In contrast, for fields applied parallel to the $c$-axis, $H\parallel c$, the magnetization remains small, nearly linear, and unsaturated up to $\mu_0H=14$~T over the measured temperature range. These data establish the $c$-axis as a hard magnetic direction and show that the low-energy Dy$^{3+}$ moments are confined predominantly to the crystallographic basal plane. We avoid referring to this as a purely ``easy-plane'' anisotropy, however, because previous single-crystal and neutron studies of DyOCl report ordered moments selected along the crystallographic $a$-axis, and recent work on DyOBr likewise finds strong in-plane anisotropy, antiferromagnetic order, and field-induced metamagnetic or plateau-like behavior for in-plane fields \cite{pan_magnetic_2024,lin_antiferromagnetic_2024}. Thus, the appropriate description is strong hard-$c$-axis anisotropy together with an additional in-plane anisotropy that becomes important in the ordered state.

At low temperatures, $T\lesssim 5$~K, the low-field response below $\mu_0H\approx 2$~T is initially similar for $H\!\perp\!c$ and $H\parallel c$ in DyOCl, and only weakly different in DyOBr. For DyOCl [Fig.~\ref{fig:mag:isomagcl}], the two field orientations separate near $\mu_0H\approx 1.8$~T, consistent with the metamagnetic feature observed in powder measurements and with the spin-flip transition reported previously for fields applied along the in-plane easy direction~\cite{tian_dyocl_2021}. For $H\!\perp\!c$, an intermediate plateau-like regime persists until $\mu_0H\approx 2.4$~T, above which the magnetization increases rapidly toward saturation. DyOBr displays qualitatively similar behavior, with somewhat different characteristic fields [Fig.~\ref{fig:mag:isomagbr}], consistent with recent single-crystal measurements that reported strong in-plane anisotropy and field-induced plateau behavior in this compound~\cite{pan_magnetic_2024}. The metamagnetic transitions and rapid saturation for $H\!\perp\!c$, together with the weak response for $H\!\parallel\!c$, demonstrate that the dominant single-ion anisotropy confines the Dy moments to the basal plane, while the ordered phases likely select specific in-plane directions through crystal-field and/or exchange anisotropy.

\subsection{\label{results:order} Thermodynamic Properties and Phase Transitions}

\begin{figure}[t]
    \centering
    \includegraphics[width=.45\textwidth]{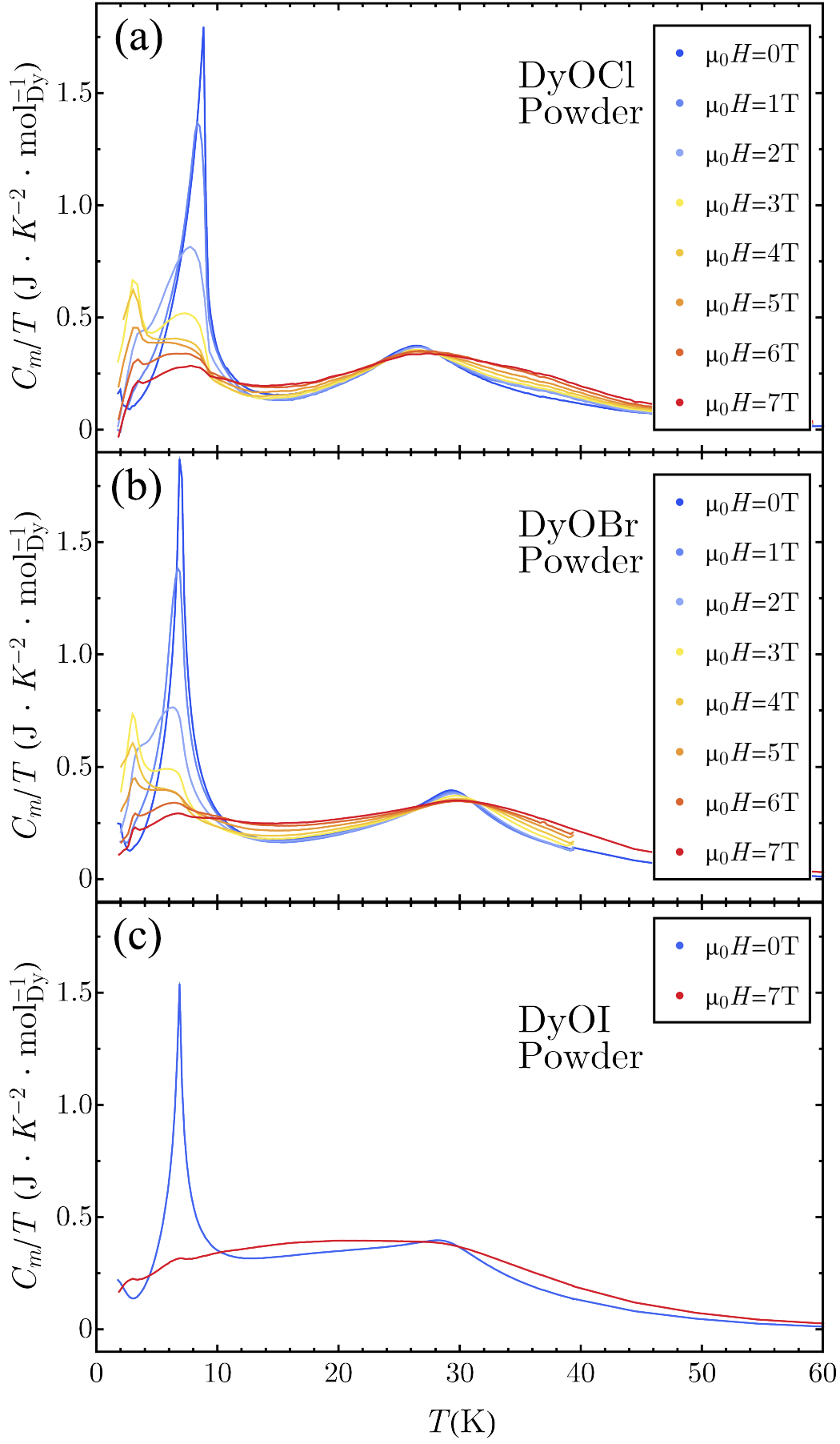}
    \caption{Molar specific heat as a function of temperature and fixed applied magnetic field for powder samples of DyOCl, DyOBr and DyOI.
    \label{fig:hc:pow}}
\end{figure}
\begin{table}[b]
        \begin{ruledtabular}
        \begin{tabular}{lccc}
        \textrm{Compound}&
        \textrm{$\theta_D$ (K)}&
        \multicolumn{1}{c}{\textrm{$T_{\rm N}$ (K)}}&
        \textrm{$T_Q$} (K)\\
        \colrule
        DyOCl & 308 & 8.84 & 26.5\\
        DyOBr & 227 \& 1170  & 6.95 & 29.9\\
        DyOI & 197 \& 860 & 6.85 & 28.6\\
        \end{tabular}
        \end{ruledtabular}
    \caption{Analysis of results from specific heat measurements on DyO$X$ powder samples. All ordering temperatures are reported inzero magnetic field. Values are uncertain in their last digit.
    \label{tab:hc:values}}
\end{table}
\begin{figure}[t]
    \centering
    \includegraphics[width=.45\textwidth]{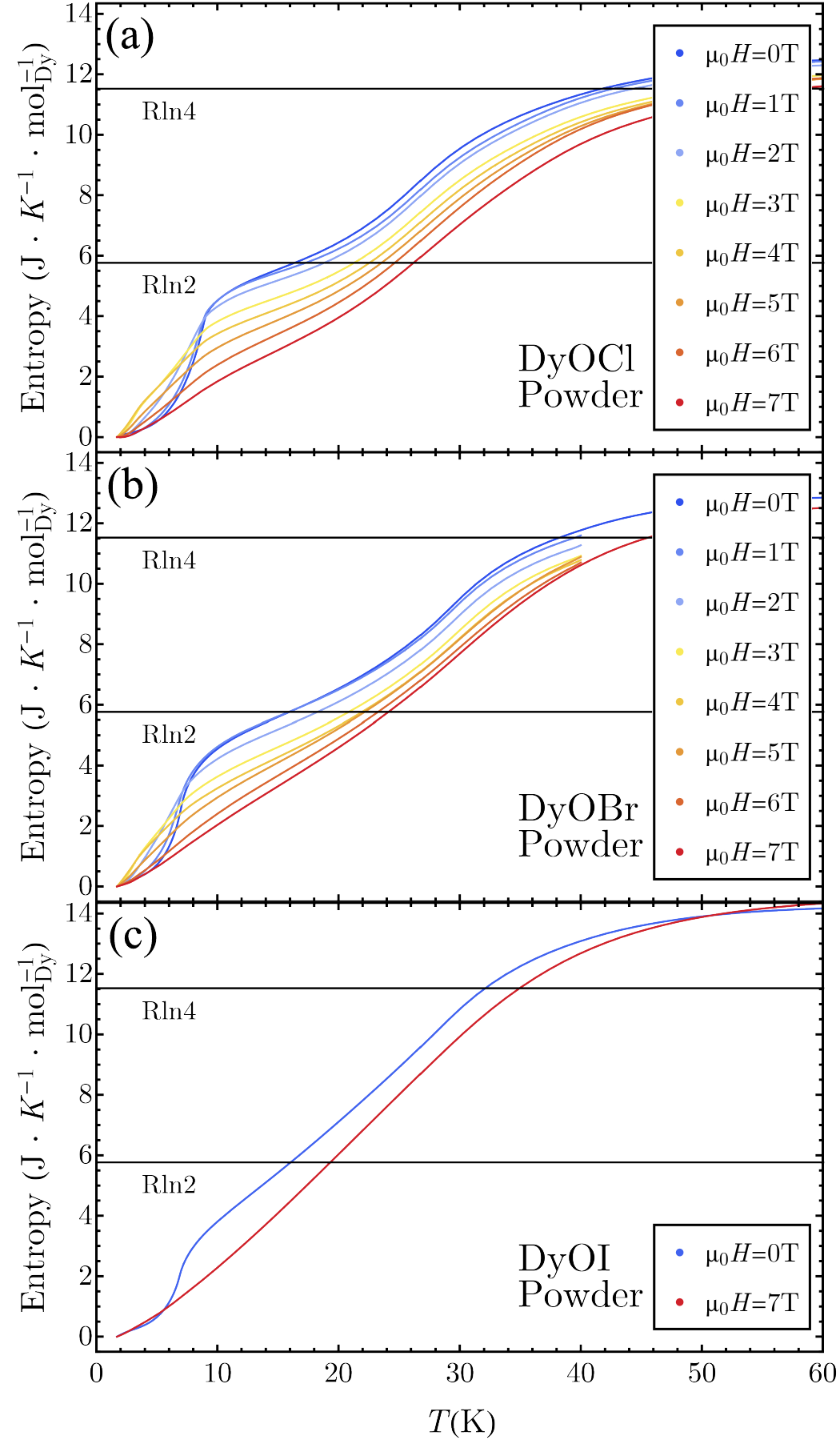}
    \caption{Entropy release as a function of temperature and applied magnetic fields for powder samples of DyO$X$. Horizontal lines correspond to entropy releases expected for two and four degrees of freedom per magnetic site, $S~=~R\ln2$ and $R\ln4$.}
    \label{fig:hc:ent}
\end{figure}

To further characterize the thermomagnetic signatures identified in Sec.~\ref{results:bulk}, we turn to heat-capacity measurements. The temperature dependence of the molar heat capacity of powder samples is shown for several applied magnetic fields in Fig.~\ref{fig:hc:pow}. All three compounds exhibit two clear low-temperature anomalies. The lower-temperature anomaly is sharp and lambda-like, occurring near $T_{\rm N}\approx 7$--$9$~K depending on composition, and is assigned to antiferromagnetic dipolar order on the basis of its corresponding susceptibility anomaly and the magnetic Bragg scattering discussed below. A second, broader anomaly occurs near $T_Q\approx 27$--$30$~K. We use the notation $T_Q$ to label this higher-temperature thermodynamic anomaly, motivated by the multipolar interpretation discussed below, but emphasize that heat capacity alone does not determine the symmetry of the associated order parameter. The transition temperatures extracted from the zero-field heat capacity are summarized in Tab.~\ref{tab:hc:values}. A susceptibility anomaly near $30$~K was also reported in single-crystal DyOBr and was tentatively attributed to crystal-field effects \cite{pan_magnetic_2024}. Our observation of a corresponding heat-capacity anomaly and substantial entropy release shows that this feature is thermodynamic, although its microscopic order parameter remains unresolved.

The two anomalies respond differently to applied magnetic field. The low-temperature antiferromagnetic anomaly is rapidly suppressed: its peak shifts to lower temperature, broadens, and loses spectral weight with increasing field. For example, by $\mu_0H=2$~T the dominant low-temperature peak has moved to $T\approx 3$~K, consistent with the metamagnetic response observed in isothermal magnetization [Fig.~\ref{fig:mag:isomagpow}]. This behavior indicates that the low-temperature anomaly is associated with the field-induced rearrangement and eventual polarization of the ordered dipolar moments. By contrast, the higher-temperature anomaly near $T_Q$ shifts only weakly with applied field, and in some field ranges moves slightly upward in temperature. This weaker field dependence suggests that $T_Q$ is not simply the onset of conventional dipolar antiferromagnetic order.

To estimate the magnetic contribution to the heat capacity, we modeled the lattice background using Debye functions fit to the high-temperature zero-field data. A single Debye temperature was used for DyOCl, while two Debye temperature scales were required to obtain satisfactory empirical descriptions of DyOBr and DyOI. The latter should be regarded as a phenomenological lattice subtraction rather than a microscopic phonon model. The need for two characteristic scales in DyOBr and DyOI may reflect their more strongly layered structures and the larger separation between intra-bilayer and inter-bilayer bonding environments. The fitted lattice contributions were subtracted from the measured heat capacities to obtain $C_{\rm m}(T)$, which was then integrated as $\Delta S(T)=\int_{T_0}^{T}\frac{C_{\rm m}(T')}{T'}\,dT'$, with $T_0=1.8$~K. The resulting entropy release is shown in Fig.~\ref{fig:hc:ent}. Because the integration begins at finite temperature and relies on a modeled lattice background, the absolute entropy values should be interpreted with the corresponding systematic uncertainty.

For all three compounds, the magnetic entropy release is consistent with two distinct thermodynamic regimes. The entropy accumulated through the low-temperature transition is of order $R\ln 2$, as expected for the ordering of a Kramers doublet. By $T=60$~K, the total entropy approaches approximately $R\ln 4$, indicating that additional low-energy degrees of freedom beyond a single isolated doublet participate in the thermodynamics. A natural interpretation is that low-lying crystal-field states generate an effective quasi-quartet manifold, allowing multipolar degrees of freedom to become relevant near $T_Q$. This scenario is consistent with quadrupolar and magnetoelastic phenomena reported in other Dy-based materials~\cite{zaharko_quadrupolar_2004,okuyama_quadrupolar_2005,watanuki_geometrical_2005,yasui_investigation_2009,popova_high-resolution_2017,nakamura_quadrupole-strain_1994,ye_elastocaloric_2022}. However, the present thermodynamic data do not by themselves establish long-range quadrupolar order. Alternative possibilities, including crystal-field population effects, magnetoelastic coupling, and the development of short-range correlations, must also be considered. The entropy evolution also shows subtle differences across the series. DyOCl and DyOBr display similar entropy releases below $T\approx 15$~K, consistent with well-separated low- and high-temperature anomalies. DyOI shows a less sharply separated entropy evolution: the entropy remains below $R\ln 2$ near $T\approx 10$~K and reaches a somewhat larger value by $T=60$~K. This suggests that the separation between the antiferromagnetic transition and the higher-temperature anomaly is less clean in DyOI, possibly reflecting stronger overlap between magnetic correlations, crystal-field population effects, and the increased two-dimensional character of the iodide compound.

\begin{figure}[t]
    \centering
    \includegraphics[width=.5\textwidth]{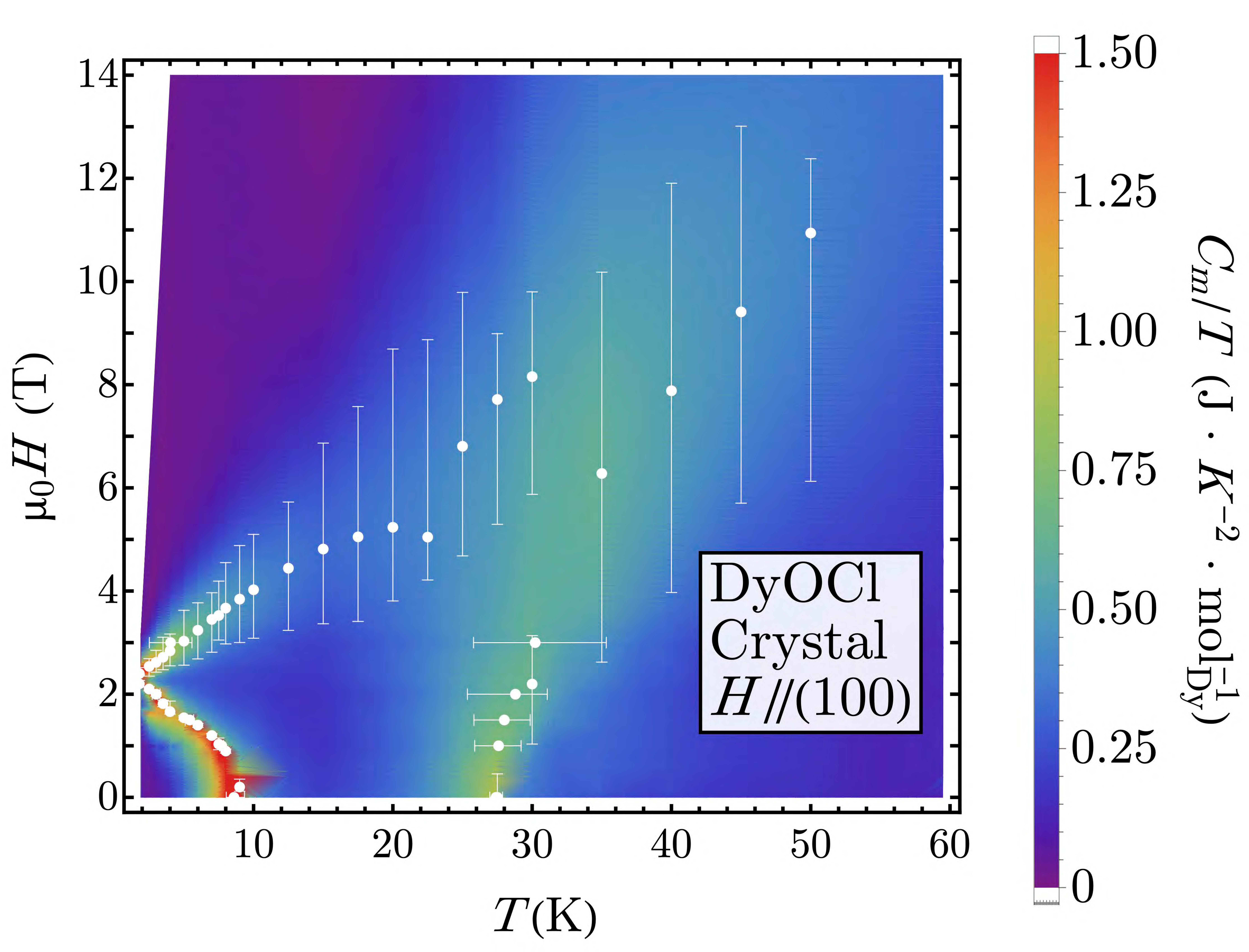}
    \caption{Interpolated color map of the temperature and magnetic-field dependence of the molar heat capacity of DyOCl. The magnetic field was applied along the crystallographic $(100)$ direction.}
    \label{fig:hc:xtal}
\end{figure}

To further examine the interplay between $T_{\rm N}$, $T_Q$, and magnetic field, we measured the heat capacity of a DyOCl single crystal with the field applied along the crystallographic $(110)$ direction. Figure~\ref{fig:hc:xtal} shows an interpolated color plot of the magnetic heat capacity as a function of temperature and field. Local maxima in the heat capacity are identified from zeros of ${\rm d}C_{\rm m}/{\rm d}T$, with uncertainty intervals estimated from nearby zeros of ${\rm d}^2C_{\rm m}/{\rm d}T^2$. As in the powder analysis, a Debye lattice contribution was subtracted before constructing the phase diagram. The resulting phase diagram contains four regimes. At low temperature and low field, DyOCl is in the antiferromagnetically ordered dipolar phase. At high temperature and low field, the system is paramagnetic. At low temperature and high field, the magnetization data indicate a field-polarized regime. Between these limits lies a regime bounded by the higher-temperature anomaly near $T_Q$, which we associate with a candidate multipolar regime. The antiferromagnetic transition moves rapidly to lower temperature with increasing field, while the higher-temperature anomaly is much less strongly suppressed. At high field and intermediate temperature, the heat-capacity features associated with the candidate multipolar regime and the field-polarized regime approach one another and become difficult to distinguish.

Taken together, the heat-capacity data establish that DyO$X$ compounds host two thermodynamic anomalies: a conventional dipolar antiferromagnetic transition at $T_{\rm N}$ and a higher-temperature anomaly near $T_Q$ involving additional low-energy degrees of freedom. The entropy balance and weak field dependence of $T_Q$ are consistent with candidate multipolar physics, but direct quadrupolar-sensitive probes would be required to identify the order parameter. We therefore next examine neutron diffraction and inelastic neutron scattering to determine which features of the thermodynamics are visible in the dipolar magnetic structure and in the crystal-field spectrum.


\subsection{\label{results:dipole} Dipolar Magnetic Order}

\begin{figure}[t]
    \centering
    \includegraphics[width=.5\textwidth]{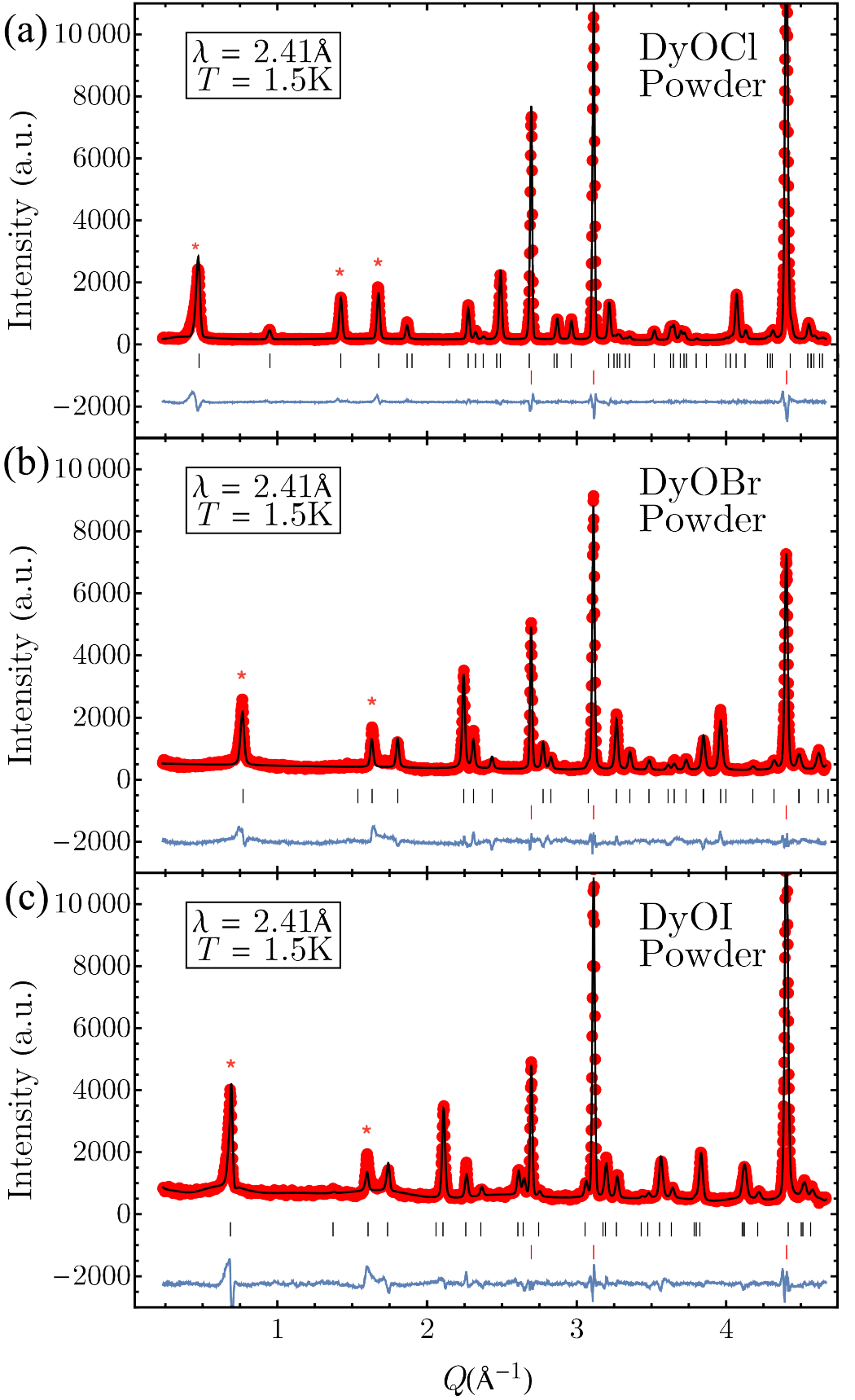}
    \caption{Powder neutron diffraction patterns (red points) for (a) DyOCl, (b) DyOBr, and (c) DyOI measured with $\lambda=2.41$~\AA\ at $T=1.5$~K. The black curves are the best-fit magnetic Rietveld refinements, and the blue curves show the corresponding data-minus-model differences. Nuclear and magnetic reflection positions of the sample are marked by black ticks, while reflections from the aluminum sample environment are marked by red ticks. The principal magnetic Bragg reflections are marked by asterisks. The low-angle profiles of DyOBr and DyOI contain broad and asymmetric magnetic scattering that is not captured by the long-range-ordered models.}
    \label{fig:struct:orderrefinement}
\end{figure}

For all three compounds, additional magnetic scattering is observed in the $T=1.5$~K neutron powder diffraction data [principal magnetic reflections marked $\ast$ in Fig.~\ref{fig:struct:orderrefinement}], whereas no additional Bragg intensity is resolved at intermediate temperatures above $T_{\rm N}$ [Fig.~\ref{fig:app:NPDdiff}]. Candidate magnetic structures were generated using the Bilbao Crystallographic Server~\cite{aroyo_bilbao_2006,aroyo_bilbao_2006-1,aroyo_crystallography_2011}. The resulting refinements are shown in Fig.~\ref{fig:struct:orderrefinement}.

For DyOCl, the magnetic scattering is well described by the propagation vector $\mathbf{k}_m=(0,0,1/2)$ and the magnetic space group $C_c mca$. The refined ordered moments lie along the crystallographic $a$ axis and have a magnitude of $7.52\,\mu_{\rm B}$/Dy. The structure consists of ferromagnetically aligned moments within each Dy monolayer, antiferromagnetic alignment of the two monolayers within a bilayer, and ferromagnetic alignment across the van der Walls gap, producing a doubling of the crystallographic unit cell along $c$ [see Fig.~\ref{fig:struct}(a) for a sketch of the magnetic structure]. The in-plane moment direction, magnetic stacking sequence, and doubled periodicity are consistent with the $A$-type antiferromagnetic structure previously reported for DyOCl~\cite{tian_dyocl_2021}. Our refined ordered moment is smaller than the value of approximately $10\,\mu_{\rm B}$/Dy reported in that study. Considering the strong neutron absorption of natural Dy and the correlations among the absorption correction, scale factor, and magnetic moment in a powder refinement, we regard the moment direction and magnetic periodicity as more robust than the absolute refined moment amplitude.

The magnetic diffraction patterns of DyOBr and DyOI are qualitatively different from that of DyOCl, both in the positions and intrinsic widths of their magnetic reflections [Fig.~\ref{fig:struct:orderrefinement}(b,c)]. The sharp component can be indexed with the propagation vector $\mathbf{k}_m=(0,0,0)$, and its intensity is reasonably reproduced by the magnetic space group $P4/n'm'm'$. Within this periodic description, however, the moments are directed along $c$ and refine to $3.54\,\mu_{\rm B}$/Dy for DyOBr and $2.57\,\mu_{\rm B}$/Dy for DyOI. These refinements should not be interpreted as microscopic structures because, in both compounds, the low-angle magnetic scattering contains broad and asymmetric components that cannot be represented by a three-dimensionally periodic magnetic structure. Moreover, the inferred $c$-axis moment direction directly conflicts with the strong hard-$c$-axis anisotropy established independently by single-crystal magnetization and the crystal-electric-field analysis described below. The $P4/n'm'm'$ refinements therefore provide an indexing and parameterization of the sharp component of the scattering rather than a self-consistent description of the complete magnetic state.

To isolate this sharp component, the DyOBr and DyOI refinements were restricted to $Q>2.0$~\AA ($2\theta\!>\!45\degree$). Peak profiles were modeled using a Thompson-Cox-Hastings pseudo-Voigt function with axial-divergence asymmetry, and the aluminum reflections were refined independently. Low-order polynomial backgrounds were sufficient for DyOCl and DyOBr, whereas a higher-order polynomial was required for DyOI to account empirically for the more pronounced diffuse contribution. The substantially better agreement obtained for DyOCl is consistent with well-developed three-dimensional magnetic order in that compound. By contrast, the coexistence of sharp and diffuse magnetic scattering in DyOBr and DyOI points to strong correlations within the Dy bilayers but reduced coherence in their relative stacking.

The origin of the apparently unphysical $c$-axis refinement can be understood from the low-order magnetic reflections. Models with either in-plane or out-of-plane moments can produce the strong $(010)$ magnetic reflection. Within a strictly periodic description, however, the in-plane models also produce sharp $(001)$ and $(002)$ reflections that are not observed as distinct peaks in the data [Fig.~\ref{fig:struct:orderrefinement}(b,c)]. A refinement constrained to the sharp scattering can therefore favor
the $c$-axis configuration. This conclusion can be misleading if the inter-bilayer registry has only finite coherence because the magnetic intensity associated with $(001)$ and $(002)$ is broadened but not absent.

The resulting in-plane interpretation is also consistent with previous studies of DyOBr, which report strong in-plane anisotropy and moments preferentially oriented along the crystallographic $a$ axis ~\cite{pan_magnetic_2024,lin_antiferromagnetic_2024}.

\begin{figure}[t]
    \centering
    \includegraphics[width=.5\textwidth]{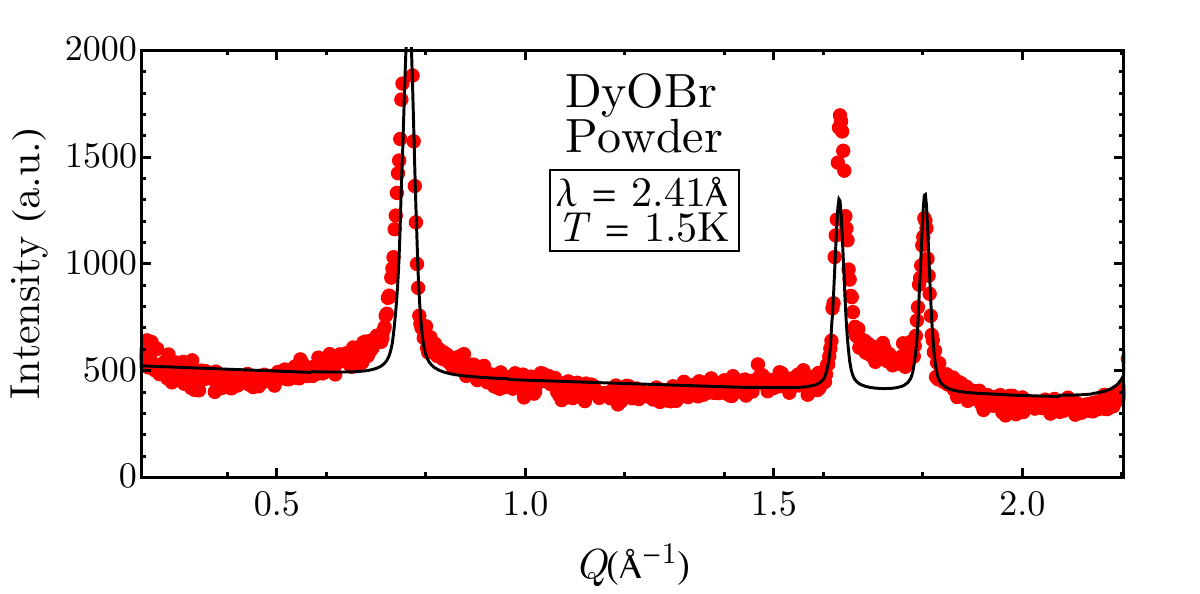}
    \caption{Low-angle portion of the $T=1.5$~K DyOBr diffraction pattern and the long-range-ordered refinement from Fig.~\ref{fig:struct:orderrefinement}. The model reproduces the principal sharp reflections but not the broad magnetic component, including the asymmetric Warren-like profile. This mismatch motivates a description with strong in-plane correlations and finite inter-bilayer coherence.}
    \label{fig:struct:dyobrzoomed}
\end{figure}

Figure~\ref{fig:struct:dyobrzoomed} highlights the broad component in DyOBr. Its asymmetric, Warren-like profile is characteristic of scattering from correlations that are extended in two dimensions but limited along the stacking direction~\cite{warren_x-ray_1941}. The profile alone does not uniquely determine a microscopic spin configuration, especially in a powder measurement, but it strongly suggests that the Dy bilayers are internally well correlated while their relative registry along $c$ remains imperfect.

To test this interpretation, we modeled the magnetic difference scattering using reverse Monte Carlo refinements implemented in \texttt{SPINVERT}~\cite{paddison_spinvert_2013}. The simulations constrain classical moments to the crystallographic $ab$ plane and allow their correlations to vary along $c$. The model supercell contains 1000 crystallographic unit cells along the stacking direction and is repeated periodically in plane. Fits with moments along any fixed in-plane direction are of comparable quality in powder diffraction; we show the $a$-axis solution because it connects continuously to the ordered structure of DyOCl. This directional choice should not be interpreted as an independent determination of the in-plane easy axis in DyOBr or DyOI.

\begin{figure}[t]
    \centering
    \includegraphics[width=.5\textwidth]{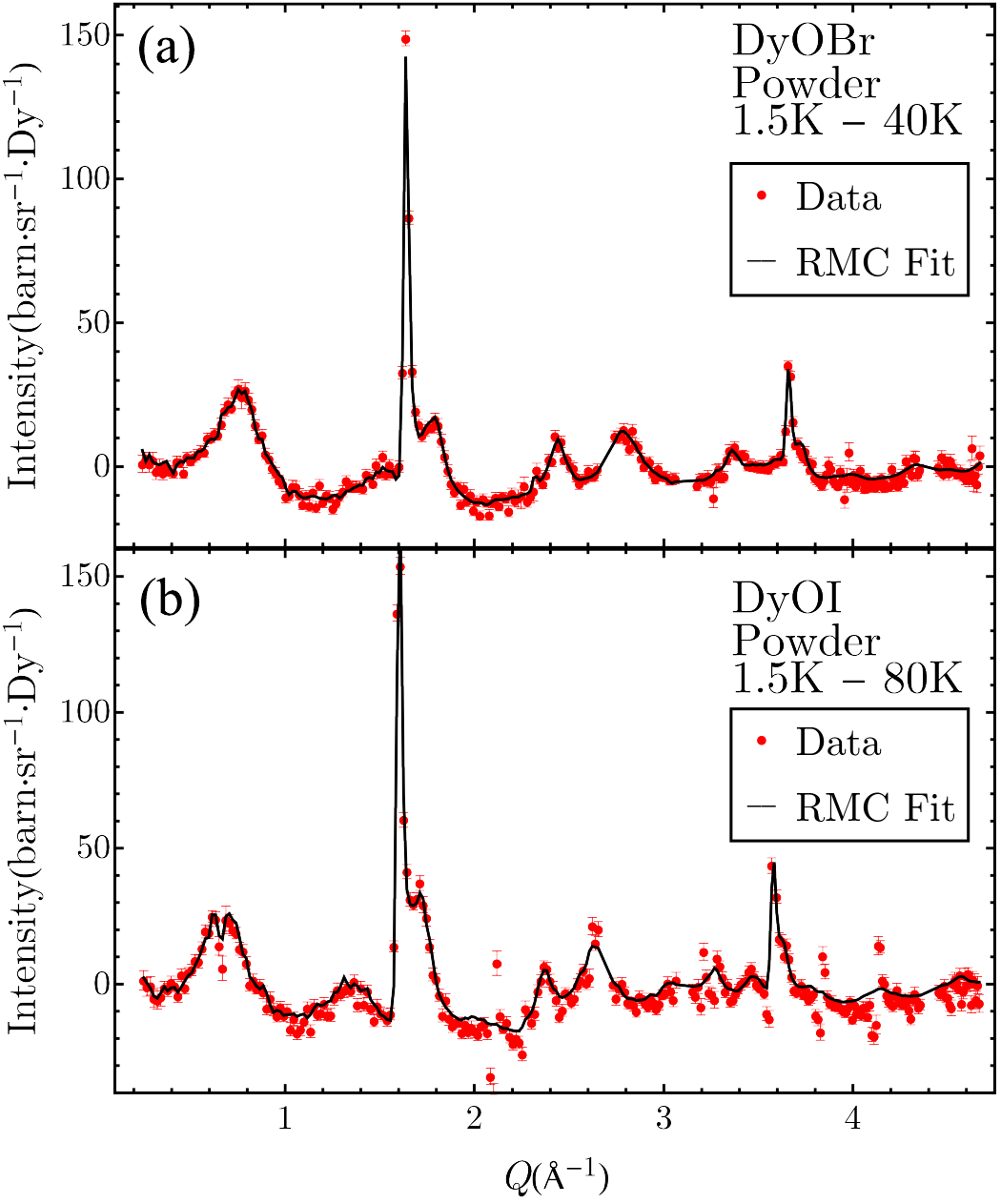}
    \caption{Reverse Monte Carlo fits to the magnetic difference scattering of (a) DyOBr and (b) DyOI using in-plane moments with finite coherence of the inter-bilayer stacking sequence. The model reproduces both the sharp magnetic reflections and the broad low-angle component. Small derivative-like features arise from subtraction of datasets collected at different temperatures and therefore slightly different lattice parameters.}
    \label{fig:struct:dyobrifit}
\end{figure}

As shown in Fig.~\ref{fig:struct:dyobrifit}, this model captures both the sharp reflections and the broad diffuse profiles of DyOBr and DyOI substantially better than a purely periodic magnetic structure. The fitted data are low-temperature minus high-temperature differences. This procedure removes the dominant nuclear scattering but also produces small derivative-like residuals because of thermal contraction between the two measurements, most visibly for DyOI. Absolute moment amplitudes extracted from such difference fits are correlated with the scale factor and background subtraction. Their numerical values are therefore less reliable than the correlation pattern, which is the principal result of the analysis.

\begin{figure}[t]
    \centering
    \includegraphics[width=.5\textwidth]{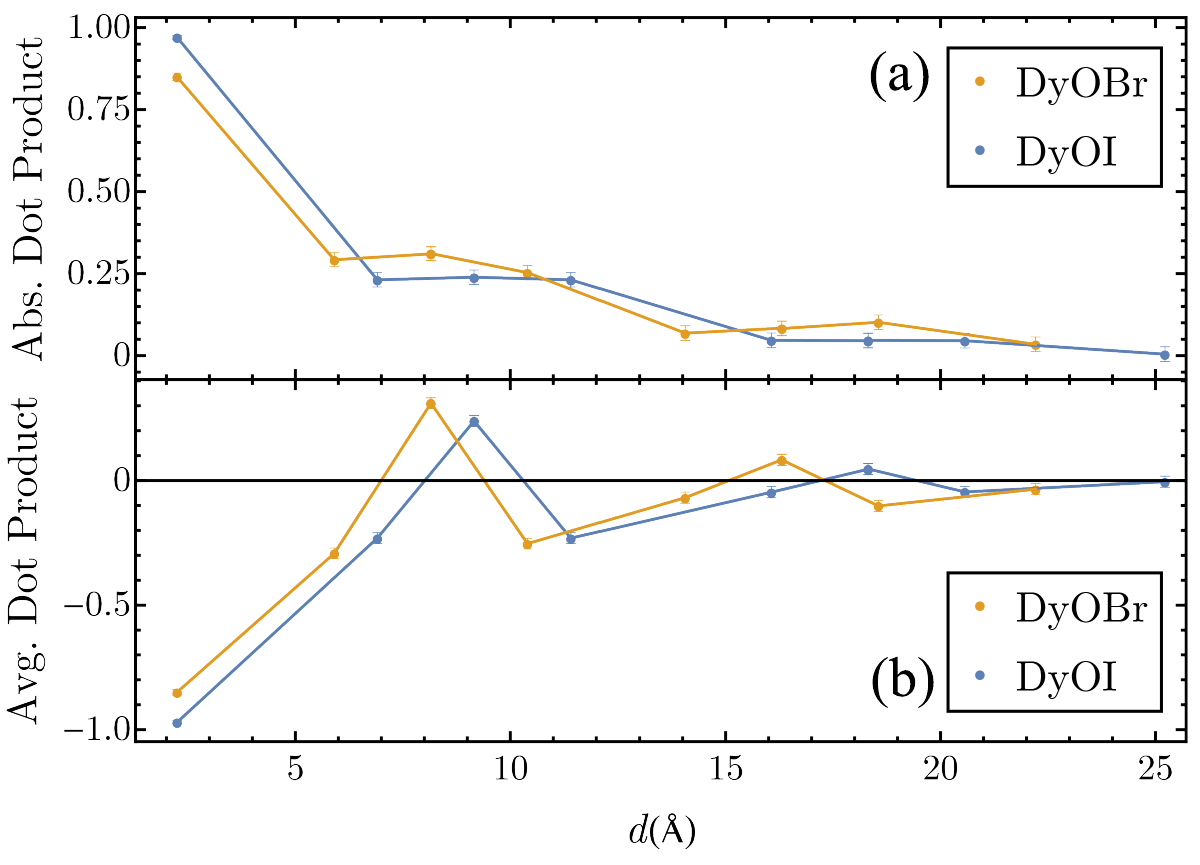}
    \caption{Layer-resolved spin correlations obtained from the reverse Monte Carlo refinements of DyOBr and DyOI. The lower panel shows the average normalized dot product between Dy monolayers as a function of their separation along $c$; the upper panel shows its magnitude. Solid lines are guides to the eye. Strong antiferromagnetic correlations occur within a bilayer and persist across the nearest van der Waals gap, but their magnitude decays with increasing layer separation.}
    \label{fig:spinvert:dyobricorrelations}
\end{figure}

The real-space correlations are summarized in Fig.~\ref{fig:spinvert:dyobricorrelations}. The shortest separation corresponds to the two Dy monolayers within one bilayer and is strongly antiferromagnetic. The next separation connects monolayers across the nearest van der Waals gap and is also antiferromagnetic on average. Correlations then decay with increasing layer separation, producing the coexistence of a sharp component and diffuse scattering. In this sense, DyOBr and DyOI retain strong intra-bilayer correlations but possess only finite inter-bilayer phase coherence. This differs from DyOCl, for which ferromagnetic alignment across the van der Waals gap and the resulting alternation of bilayers doubles the magnetic unit cell along $c$. 

A model in which the disorder is confined to the stacking direction accounts simultaneously for three features of the magnetic diffraction: sharp reflections, broad diffuse scattering, and an asymmetric Warren-like profile. The sharp reflections originate from the well-correlated magnetic structure in the $ab$ plane. Reflections with a substantial $L$ component, including $(001)$ and $(002)$, are broadened by the finite coherence of the inter-bilayer registry. Reflections containing both in-plane and out-of-plane components produce an asymmetric Warren-like powder profile as the corresponding scattering rods intersect the Ewald sphere. By contrast, broadened $(00L)$ scattering does not develop the same characteristic asymmetry. The absence of sharp $(001)$ and $(002)$ reflections should therefore not be interpreted as evidence against predominantly in-plane moments. In this way, a model with stacking-fault-like disorder captures several otherwise disparate features in a self-consistent manner. 

The correlation profiles derived from powder reverse Monte Carlo refinements are not unique microscopic structures. They demonstrate that an in-plane-moment model with imperfect inter-bilayer registry is consistent with the principal features of the magnetic scattering and resolves the contradictions of the long-range refinements. Distinguishing random stacking faults from finite correlation lengths, competing stacking domains, or more complex modulations will require single-crystal diffuse scattering.

\subsection{\label{results:cef} Crystal-Electric-Field and Magnetic Excitations}

\begin{figure}[t]
    \centering
    \includegraphics[width=.5\textwidth]{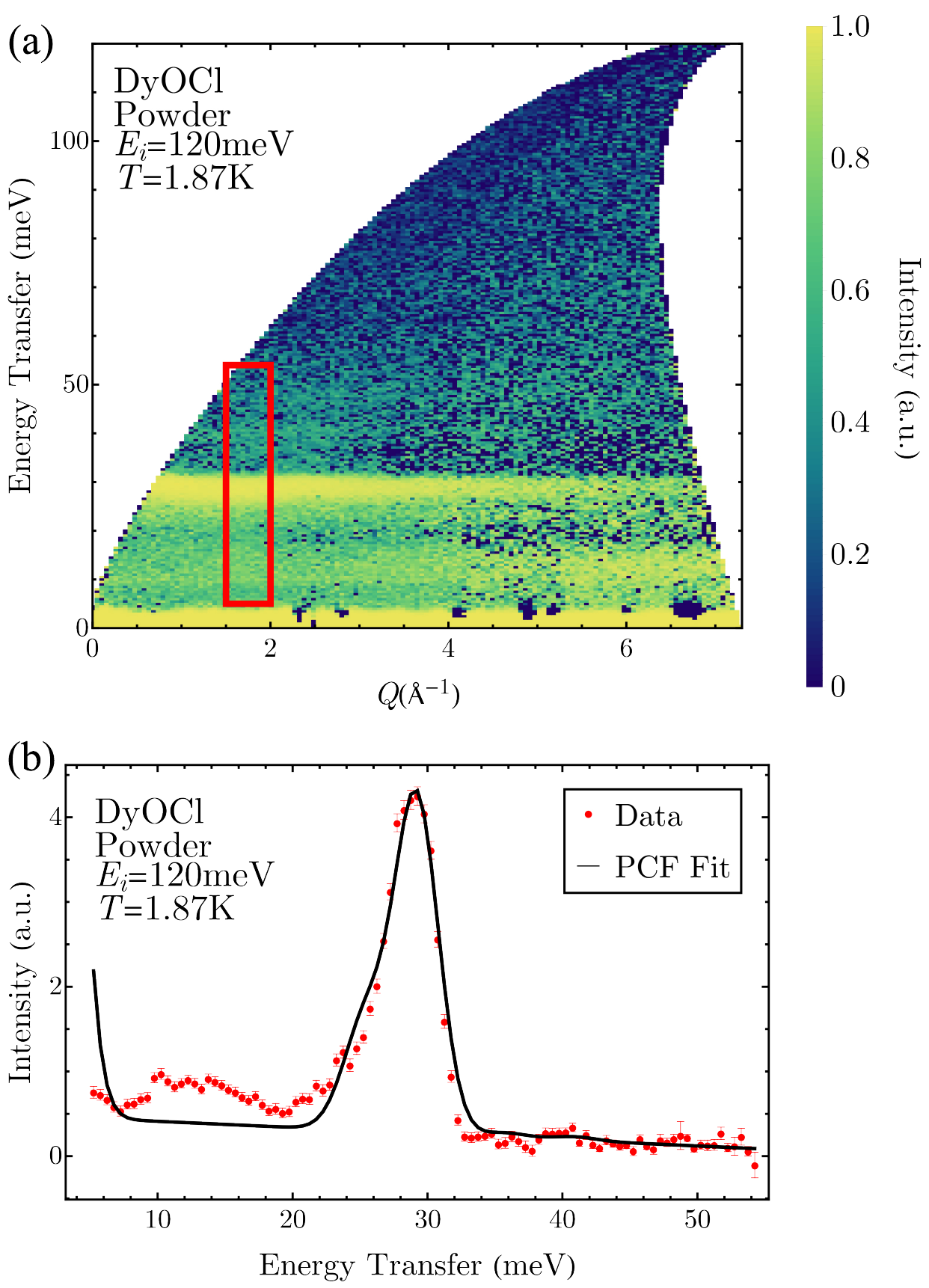}
    \caption{Inelastic neutron scattering from DyOCl measured on SEQUOIA with $E_i=120$~meV. (a) Powder-averaged intensity as a function of momentum transfer $Q$ and energy transfer $E$. The modes near 24 and 29~meV decrease with increasing $Q$ and are assigned to crystal-electric-field excitations, whereas the broader intensity extending from approximately 10 to 30~meV increases with $Q$ and is predominantly from phonons. (b) Constant-$Q$ cut integrated over the region indicated in panel (a), together with the best-fit point-charge crystal-field model.}
    \label{fig:cf:seq_data}
\end{figure}

We now turn to magnetic excitations in DyOCl. Fig.~\ref{fig:cf:seq_data}(a) shows the inelastic neutron scattering $\tilde{I}(Q,E)$ from DyOCl measured on SEQUOIA. Two overlapping excitations are resolved near 24 and 29~meV, with fitted Gaussian full widths at half maximum of 3.8 and 4.0~meV, respectively [Fig.~\ref{fig:cf:seq_data}(b)]. Their intensity decreases with increasing $Q$, as expected for magnetic scattering from localized $4f$ moments. A broader band extending over approximately 10--30~meV instead gains intensity with $Q$ and is therefore assigned predominantly to phonon scattering. The constant-$Q$ cut  was integrated over $1.5\leq Q\leq2.0$~\AA$^{-1}$.

We modeled the magnetic excitations using \texttt{PyCrystalField}~\cite{scheie_pycrystalfield_2021}. The starting Hamiltonian was obtained from a point-charge calculation using the ligand coordinates determined by neutron diffraction. After subtraction of a linear background, the effective oxygen and chlorine charges were allowed to vary through multiplicative renormalization factors, and the calculated neutron spectrum was convolved with the instrumental resolution and scaled to the data. A broad family of charge combinations produces nearly equivalent descriptions of the two observed excitations. To avoid overparameterization, we retain the solution in which the oxygen and chlorine charges are renormalized uniformly. The optimal scale factor is 1.196 and gives the Stevens parameters $B_2^0=0.50$, $B_4^0=7.7\times10^{-4}$, $B_4^4=-6.0\times10^{-3}$, $B_6^0=2.6\times10^{-6}$, and $B_6^4=2.6\times10^{-5}$~meV, in the conventions used by \texttt{PyCrystalField} [See Tab.~\ref{tab:cffit} for wavefunctions]. The corresponding model reproduces the unresolved structure of the 24 and 29~meV peaks. Because the available powder data constrain only a limited number of transitions, these parameters should be regarded as an effective crystal-field model rather than a unique determination of the microscopic ligand charges.

The calculated ground-state doublet has a strongly anisotropic effective $g$ tensor, with $g_{xx}=g_{yy}\approx21.0$ and $g_{zz}\approx0.96$. These values correspond to saturation moments of approximately $10.5\,\mu_{\rm B}$/Dy in the basal plane and $0.48\,\mu_{\rm B}$/Dy along the $c$ axis, consistent with the strong hard-$c$-axis response observed in the single-crystal magnetization. The same model predicts a neutron-active transition from the ground-state Kramers doublet to a second doublet near $0.73$~meV. Because this energy is below the useful resolution of the SEQUOIA measurement, dedicated low-energy measurements were performed to search for the predicted excitation.

\begin{figure}[t]
    \centering
    \includegraphics[width=.5\textwidth]{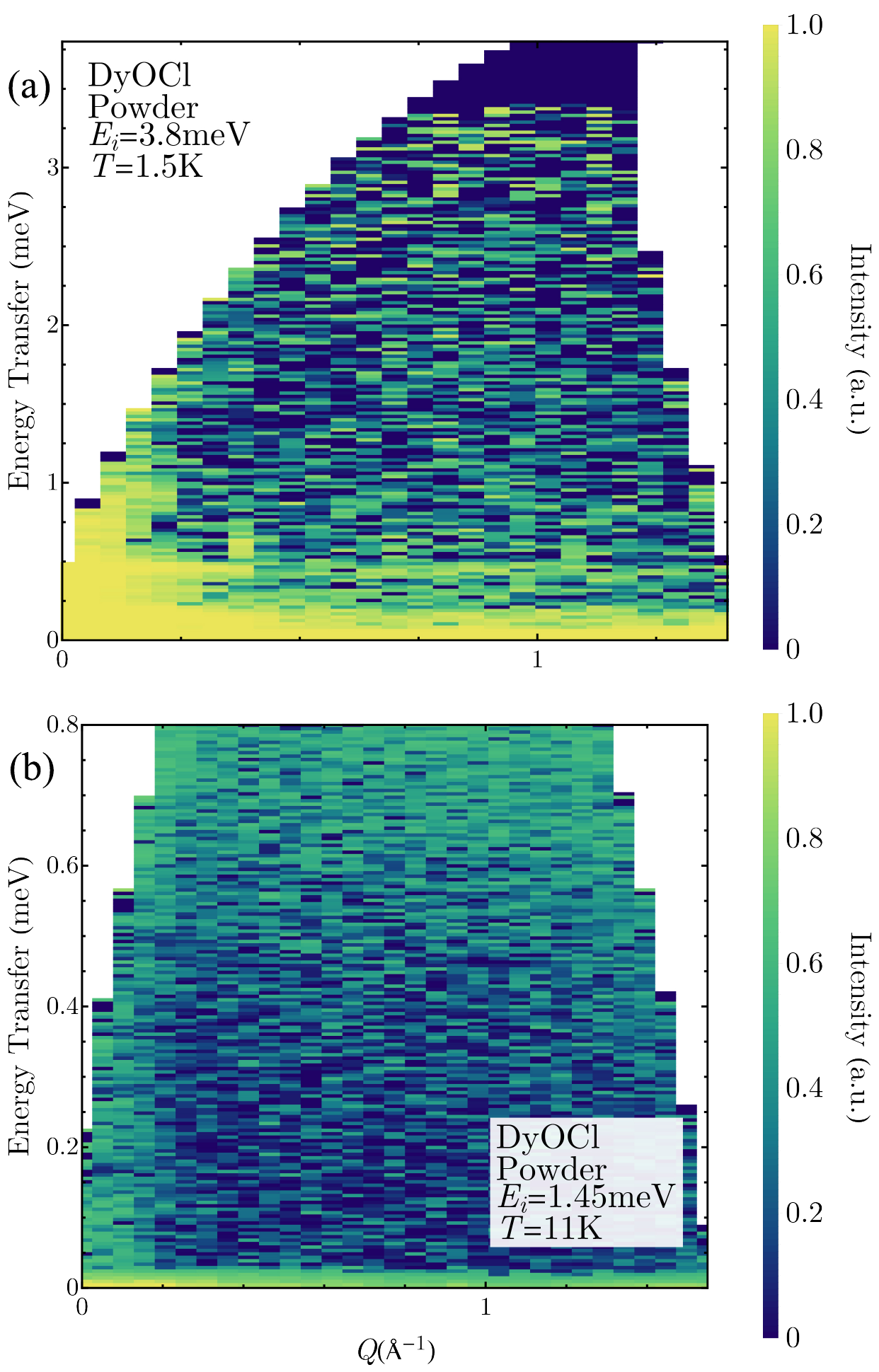}
    \caption{Low-energy inelastic neutron scattering from DyOCl measured on (a) HYSPEC and (b) DCS. No reproducible magnetic excitation is resolved near the $0.3$--$0.8$~meV energy range predicted by the point-charge crystal-field models. The feature near $E\!=\!0.5$~meV in panel (a) is an instrumental artifact observed for this measurement geometry.}
    \label{fig:cf:0p8}
\end{figure}
\begin{figure*}
\includegraphics[width=\textwidth]{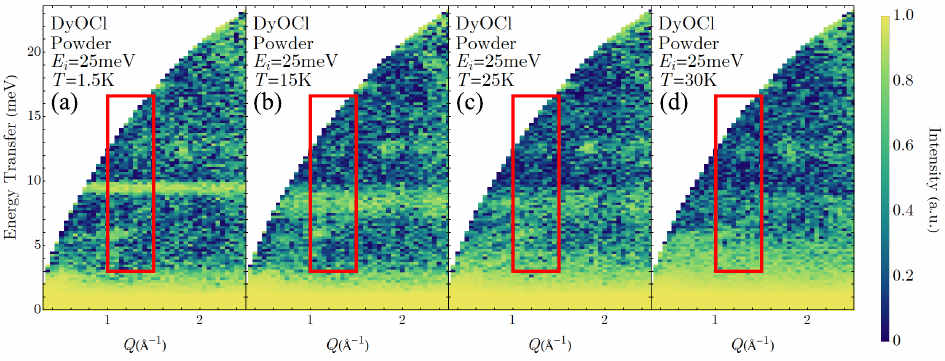}
\caption{Temperature dependence of the 10~meV excitation in DyOCl measured on HYSPEC with $E_i=25$~meV. Panels (a)--(d) show the powder-averaged scattering at the indicated temperatures. The red rectangles mark the integration range used to obtain the constant-$Q$ cuts in Fig.~\ref{fig:cf:10meVdep}. The mode broadens, shifts to lower energy, and loses spectral weight upon warming toward the anomaly at $T_Q$.}
\label{fig:cf:hys_data}
\end{figure*}
\begin{figure}
\includegraphics[width=0.5\textwidth]{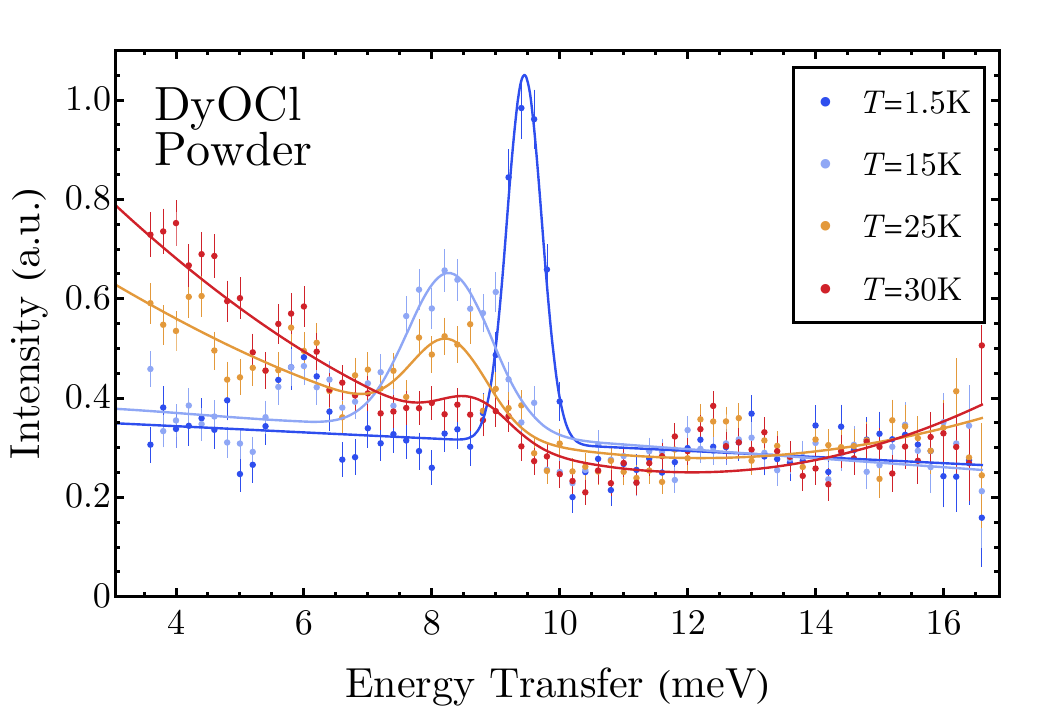}
\caption{Constant-$Q$ cuts through the 10~meV excitation of DyOCl, integrated over the regions marked in Fig.~\ref{fig:cf:hys_data}. Solid curves are fits consisting of a Gaussian peak and a polynomial background. The excitation broadens and softens with increasing temperature and is no longer resolved near $T_Q$.}
\label{fig:cf:10meVdep}
\end{figure}

The low-energy HYSPEC and DCS spectra are shown in Fig.~\ref{fig:cf:0p8}. Neither experiment resolves a reproducible magnetic mode in the $0.3$--$0.8$~meV range predicted by the crystal-field models. The feature near $E=0.5$~meV in the HYSPEC data is also present in measurements of the same instrumental configuration and is therefore assigned to an experimental artifact. Thus, although a low-lying excited doublet arises robustly within the constrained point-charge fits to the 24 and 29~meV transitions, its existence is not established experimentally. This discrepancy indicates that the high-energy powder spectrum does not constrain the low-energy crystal-field scheme uniquely, or that interactions neglected by the single-ion model substantially renormalize the low-energy states.

A separate excitation is observed near 10~meV in the HYSPEC data collected with $E_i=25$~meV [Fig.~\ref{fig:cf:hys_data}]. Over the momentum range in which it can be isolated from the elastic line and phonon background, the mode shows little measurable dispersion and has a $Q$ dependence consistent with a magnetic origin. It may contribute to the lower-energy intensity in the SEQUOIA spectrum, but there it overlaps strongly with phonon scattering. To follow its temperature evolution, we integrate the HYSPEC data over $1.0\leq Q\leq1.5$~\AA$^{-1}$, excluding the immediate vicinity of the elastic line. The resulting cuts are shown in Fig.~\ref{fig:cf:10meVdep}.

The excitation is sharpest at the lowest temperature and broadens substantially on warming. Fits using a Gaussian peak on a polynomial background give a full width at half maximum of approximately 0.62~meV at $T=1.5$~K and 1.5~meV at $T=15$~K. Its central energy also shifts downward as $T_Q$ is approached, while the peak becomes progressively less distinct from the background. By $T=30$~K, the mode is no longer resolved as a separate inelastic feature. The evolution is therefore tied more closely to the higher-temperature thermodynamic anomaly than to the onset of dipolar order at $T_{\rm N}$. The present data establish this correlation but do not determine whether the mode is a renormalized crystal-field excitation, a collective multipolar excitation, or a mixed magnetoelastic mode.

\section{\label{sec:discussion} Discussion}

The DyO$X$ series combines a nearly unchanged square-bilayer geometry with a large and systematically tunable van der Waals separation. This structural control produces a clear contrast between common local physics and distinct collective magnetic states. All three compounds exhibit similar thermodynamic anomalies and strong hard-$c$-axis anisotropy, while their low-temperature magnetic diffraction evolves from three-dimensional long-range order in DyOCl to incomplete inter-bilayer magnetic coherence in DyOBr and DyOI. The central questions are therefore how the magnetic correlations and interactions between otherwise similar bilayers evolve across the series, and whether the higher-temperature anomaly at $T_Q$ can be connected to a specific microscopic degree of freedom.

\subsection{\label{discuss:macro} Macroscopic Phenomenology}

The bulk measurements establish two distinct thermodynamic anomalies in all three compounds. The lower-temperature anomaly at $T_{\rm N}\simeq 7$--$10$~K is sharp in heat capacity, produces a cusp in susceptibility, and coincides with the appearance of magnetic Bragg scattering. It can therefore be assigned confidently to antiferromagnetic dipolar order. The broader anomaly at $T_Q\simeq 27$--$30$~K has a different character. It carries substantial entropy, evolves only weakly with magnetic field, and is not accompanied by detectable magnetic Bragg peaks or a resolved structural transition. These observations exclude a second conventional dipolar ordering transition as the simplest explanation of $T_Q$.

A multipolar interpretation of the higher-temperature anomaly is plausible, but is not established by the present measurements. Dy$^{3+}$ is a Kramers ion, and a single isolated Kramers doublet supports only an effective pseudospin-$1/2$ degree of freedom. Independent quadrupolar degrees of freedom require the participation of additional crystal-field states, for example through a low-lying quasi-quartet. The entropy approaching approximately $R\ln 4$ by $T=60$~K is consistent with an enlarged low-energy manifold, but does not determine its level structure or identify the associated order parameter. Short-range magnetic correlations, magnetoelastic coupling, or the thermal population of crystal-field levels could produce related thermodynamic signatures. We therefore use $T_Q$ as a label for the higher-temperature anomaly and refer to the regime between $T_{\rm N}$ and $T_Q$ as a candidate multipolar regime.

The absence of a resolved crystallographic symmetry change across $T_Q$ distinguishes DyO$X$ from several Dy-based quadrupolar materials in which multipolar order is accompanied by a measurable lattice distortion~\cite{zaharko_quadrupolar_2004,watanuki_geometrical_2005, usui_observation_2014,popova_high-resolution_2017}. This absence does not exclude quadrupolar order. A distortion may lie below the resolution of the powder diffraction measurements, preserve the average crystallographic symmetry, or be unnecessary for the relevant order parameter. It does, however, show that the anomaly at $T_Q$ is not accompanied by a substantial uniform change of the tetragonal crystal structure. 

The single-crystal magnetization provides a more definitive result. In both DyOCl and DyOBr, the response for $H\perp c$ approaches the full Dy moment at moderate field, whereas the response for $H\parallel c$ remains small and nearly linear up to 14~T [Figs.~\ref{fig:mag:isomagcl} and \ref{fig:mag:isomagbr}]. Together with the crystal-field $g$ tensor, these measurements establish a strong hard-$c$-axis anisotropy. The reduced high-field magnetization of the powder samples then follows naturally from orientational averaging of this anisotropic response. The metamagnetic features observed for in-plane fields further show that the basal plane is not magnetically isotropic in the ordered state, consistent with moment selection along a crystallographic in-plane direction. 

The field-dependent heat capacity of DyOCl summarizes the separation between the two anomalies [Fig.~\ref{fig:hc:xtal}]. The phase boundary associated with $T_{\rm N}$ is rapidly suppressed by magnetic field and connects to the metamagnetic and field-polarized regimes. The feature associated with $T_Q$ evolves much more weakly. The resulting diagram contains a low-field antiferromagnetic phase, a high-field polarized regime, a paramagnetic regime, and an intermediate region bounded by the $T_Q$ anomaly. The contrasting field dependences show that the two anomalies originate from different components of the low-energy magnetic manifold. Interpreting the intermediate region as multipolar remains a physically motivated hypothesis rather than a direct identification of its order parameter.

\subsection{\label{discuss:micro} Microscopic Interpretation}

The microscopic measurements separate the two thermodynamic anomalies. Static dipolar order appears only below $T_{\rm N}$, where magnetic Bragg scattering develops in all three compounds. No additional magnetic Bragg component is resolved between $T_{\rm N}$ and $T_Q$ [Fig.~\ref{fig:app:NPDdiff}]. Within the sensitivity of the powder diffraction measurements, the state below $T_Q$ therefore carries no periodic dipolar component of appreciable magnitude. This null result is compatible with quadrupolar order, to which conventional neutron diffraction is only indirectly sensitive, but does not identify the order parameter.

The low-temperature diffraction shows that halide substitution acts most strongly on the relative organization of the Dy bilayers. The intra-bilayer Dy geometry varies little across the series, and strong antiferromagnetic correlations within each bilayer are retained. The inter-bilayer correlations instead change qualitatively. DyOCl develops a three-dimensionally ordered structure with $\mathbf{k}_m=(0,0,1/2)$ and ferromagnetic alignment across the nearest van der Waals gap. DyOBr and DyOI exhibit sharp magnetic reflections together with broad, Warren-like scattering. The reverse Monte Carlo analysis shows that these patterns are consistent with strongly correlated bilayers whose relative registry retains only finite coherence [Figs.~\ref{fig:struct:dyobrifit} and \ref{fig:spinvert:dyobricorrelations}].

The increase of the van der Waals separation provides a natural explanation for the progressive loss of stacking coherence, but does not by itself explain the change in the preferred inter-bilayer phase relation. The latter indicates that competing contributions to the inter-bilayer coupling change balance across the series. The local coordination provides one possible origin. In DyOCl, the chlorine ion across the van der Waals gap lies at a distance comparable to that of the chlorine ligands within the bilayer, whereas the corresponding separation increases rapidly for Br and I. This change can modify both the axial crystal field and the exchange pathways across the gap. Long-range dipolar interactions may also become comparatively important as shorter-range exchange interactions are reduced. A microscopic account of the changing stacking registry will therefore require the crystal-field wavefunctions, anisotropic exchange, and long-range dipolar interactions to be considered together.

The crystal-field analysis provides a partial description of the single-ion physics. The effective point-charge model reproduces the excitations near 24 and 29~meV and yields a strongly anisotropic ground-state $g$ tensor consistent with the hard-$c$-axis magnetization. Its low-energy prediction is not confirmed, however. The model produces an excited Kramers doublet near $0.73$~meV, but no corresponding transition is resolved in the HYSPEC or DCS spectra [Fig.~\ref{fig:cf:0p8}]. The entropy recovered by approximately 60~K suggests that degrees of freedom beyond one isolated Kramers doublet contribute to the thermodynamics, but the present spectroscopy does not establish the low-lying quasi-quartet that would provide the simplest local basis for independent quadrupolar moments.

The 10~meV excitation imposes a separate and more direct constraint. The mode is weakly dispersive over the measured momentum range and has a $Q$ dependence consistent with magnetic scattering. It remains visible at $T=15$~K, above $T_{\rm N}$, and therefore does not require static dipolar order. On warming toward $T_Q$, it softens, broadens strongly, and is no longer resolved as a distinct peak near the higher-temperature anomaly [Figs.~\ref{fig:cf:hys_data} and \ref{fig:cf:10meVdep}]. Its temperature dependence consequently associates it with the correlations developing below $T_Q$, although the present powder data do not determine whether it is a renormalized crystal-field excitation, a collective mode, or a mixed magnetoelastic excitation.

The energy of this mode is difficult to reconcile with the low-energy level predicted by the effective point-charge model. A direct estimate of the bare magnetostatic field generated by the ordered Dy moments, using 
\begin{equation} 
\label{eq1} \mathbf{B}(\mathbf{r}) = 
    \frac{\mu_0}{4\pi r^3} \left[ 3\hat{\mathbf{r}} \left( \hat{\mathbf{r}}\cdot\boldsymbol{\mu} \right) 
    - \boldsymbol{\mu} \right], 
\end{equation}
gives a local field of order $0.6$~T when moments within a radius of approximately 10~\AA\ are included. This value is far below the field required in the single-ion calculation to shift the predicted sub-meV transition to approximately 10~meV. The estimate excludes a simple magnetostatic origin, but does not constrain exchange-generated molecular fields or interaction-induced hybridization between crystal-field and collective degrees of freedom.

\begin{figure}[t]
    \centering
    \includegraphics[width=0.5\textwidth]{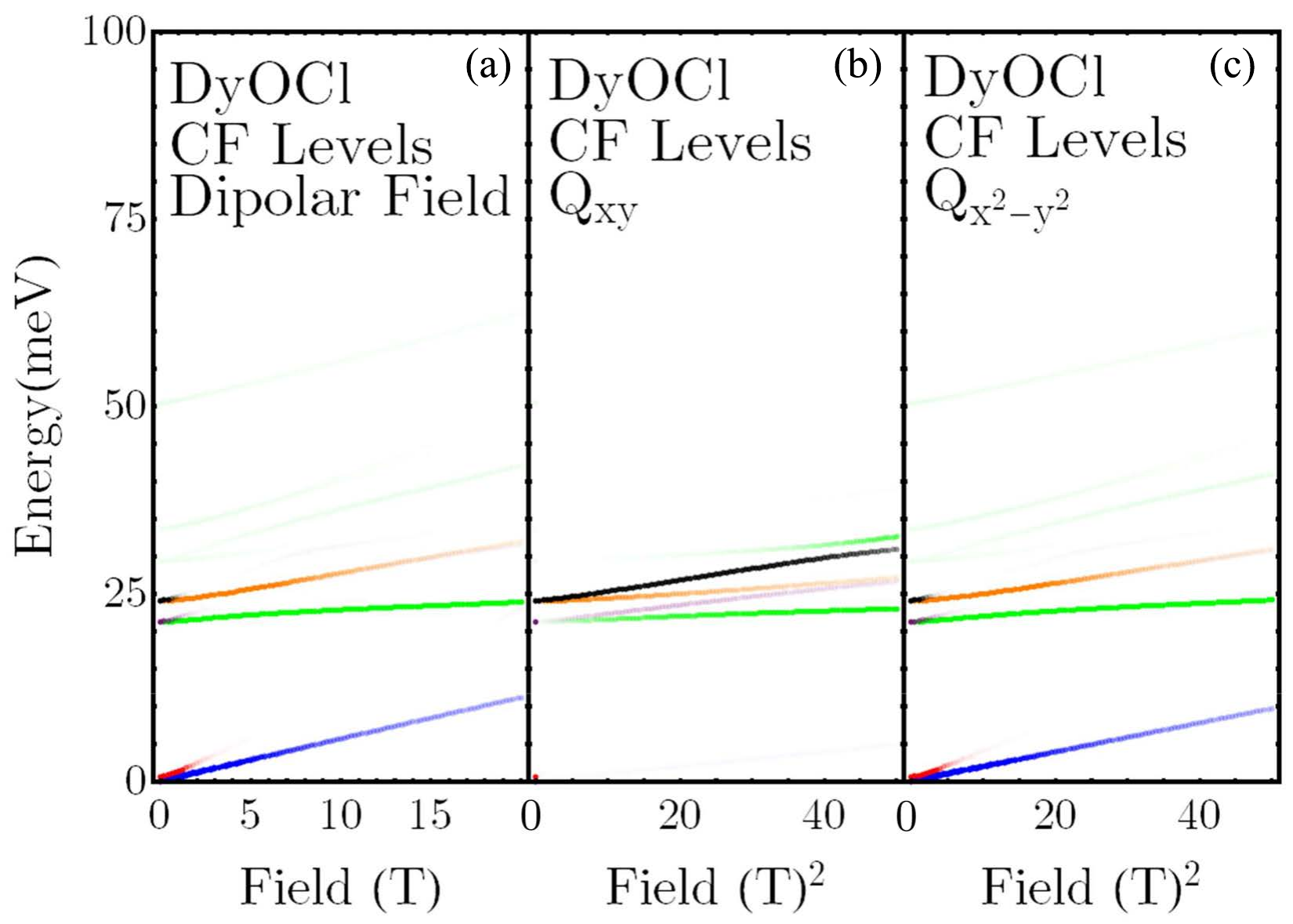}
    \caption{Calculated evolution of the DyOCl crystal-field levels under (a) a dipolar field consistent with the ordered moment direction, (b) an effective $Q_{x^2-y^2}$ field, and (c) an effective $Q_{xy}$ field. The opacity of each level is proportional to the calculated neutron-transition intensity from the thermally populated states at $T=1.8$~K. The fields can generate a level near 10~meV, but none provides a simultaneous account of the observed low- and high-energy spectra.}
    \label{fig:cf:levels}
\end{figure}

To test whether a phenomenological multipolar field could reconcile the low- and high-energy spectra, we added rank-two operators to the effective single-ion Hamiltonian. Defining the anticommutator as $\{A,B\}=AB+BA$, the Hermitian quadrupolar operators are $Q_{3z^2-r^2} =\left[3J_z^2-J(J+1)\right]/3$, $Q_{x^2-y^2} = \left(J_x^2-J_y^2\right)/\sqrt{3}$, $Q_{xy} = \{J_x,J_y\}/\sqrt{3}$, $Q_{yz} = \{J_y,J_z\}/\sqrt{3}$, and $Q_{zx} =\{J_z,J_x\}/\sqrt{3}$. Within the tested parameter range, fields coupled to $Q_{yz}$ and $Q_{zx}$ do not generate appreciable intensity in the observed neutron-active channels. The effects of $Q_{3z^2-r^2}$ and $Q_{x^2-y^2}$ are strongly correlated for the transitions considered. We therefore explored an effective parameter space containing dipolar, $Q_{x^2-y^2}$, and $Q_{xy}$ fields. This reduction is specific to the available powder spectra and should not be interpreted as a symmetry determination of the order parameter.

A grid search was performed for several uniform charge-renormalization factors, and each trial spectrum was compared simultaneously with the SEQUOIA $E_i=120$~meV and HYSPEC $E_i=25$~meV data. The calculated level evolution is summarized in Fig.~\ref{fig:cf:levels}. Dipolar and $Q_{x^2-y^2}$ fields perturb the neutron-active levels in similar ways, whereas the $Q_{xy}$ field produces a more distinct pattern. Fields of all three types can generate a neutron-active level near 10~meV. However, none of the tested models reproduces this feature together with the excitations near 25 to 30~meV without introducing additional transitions that are absent from the measurements. This negative result constrains the microscopic interpretation. The 10~meV feature cannot be described as a crystal-field transition shifted by a uniform static dipolar or quadrupolar field within the present point-charge Hamiltonian. These calculation tests a restricted class of static single-ion mean-field models and do not exclude collective multipolar excitations. In particular, an excitation of predominantly quadrupolar character could acquire dipolar neutron intensity through mixing with excited crystal-field states. Likewise, coupling between a crystal-field excitation and a phonon could produce a magnetic mode whose energy and linewidth evolve strongly through a magnetoelastic instability.

The remaining interpretations can be distinguished experimentally. A renormalized crystal-field excitation would require a low-energy level scheme different from that generated by the constrained point-charge model. A collective multipolar excitation should display a characteristic momentum dependence and acquire neutron intensity through dipolar admixture. A magnetoelastic excitation should produce corresponding anomalies in phonon, Raman, or elastic response. Polarized inelastic neutron scattering on a single crystal would separate the magnetic and nuclear components of the 10~meV feature and determine its dispersion. Resonant x-ray scattering, ultrasound or elastic-constant measurements, and Raman spectroscopy across $T_Q$ would directly test for quadrupolar order and magnetoelastic coupling. Single-crystal diffuse neutron scattering would separately determine the inter-bilayer correlation function in DyOBr and DyOI. Together, these measurements would establish whether the common anomaly at $T_Q$ and the changing inter-bilayer registry arise from distinct parts of the Dy magnetic manifold or from a coupled multipolar, dipolar, and lattice instability.

\section{\label{sec:concl} Conclusions}

We have established the dysprosium oxyhalides DyOCl, DyOBr, and DyOI as a chemically tunable family of square-bilayer rare-earth magnets. Substitution of Cl by Br and I expands the van der Waals gap by approximately 60\% while leaving the intra-bilayer Dy geometry nearly unchanged. Despite this large structural tuning, all three compounds exhibit similar bulk behavior: strong hard-$c$-axis anisotropy, antiferromagnetic Curie-Weiss interactions, a dipolar ordering transition at $T_{\rm N}\simeq7$-$10$~K, and a second thermodynamic anomaly at $T_Q\simeq27$-$30$~K.

Neutron diffraction reveals that the microscopic ordered states are considerably less uniform across the series than the thermodynamics suggest. DyOCl develops three-dimensional antiferromagnetic order with $\mathbf{k}_m=(0,0,1/2)$ and moments in the basal plane. DyOBr and DyOI instead exhibit sharp magnetic reflections coexisting with broad Warren-like scattering. Reverse Monte Carlo analysis shows that these patterns are consistent with strong in-plane and intra-bilayer correlations but finite inter-bilayer coherence. The evolution from DyOCl to DyOBr and DyOI therefore reflects a loss and change of magnetic registry across the expanding van der Waals gap, rather than a simple reduction of the ordered moment.

Inelastic neutron scattering on DyOCl resolves magnetic crystal-field excitations near 24 and 29~meV and a separate mode near 10~meV. A constrained point-charge model reproduces the high-energy pair and yields a strongly anisotropic ground-state $g$ tensor consistent with magnetization. The same model predicts a sub-meV excited doublet, but no corresponding transition is observed in high-resolution measurements. The 10~meV mode broadens, softens, and becomes unresolved on approaching $T_Q$, directly linking its spectral evolution to the higher-temperature anomaly. Static dipolar and phenomenological quadrupolar mean fields do not provide a simultaneous description of this mode and the high-energy crystal-field spectrum.

The entropy recovered through the two anomalies, the weak field dependence of $T_Q$, and the absence of dipolar Bragg order above $T_{\rm N}$ are consistent with candidate multipolar physics, but they do not establish quadrupolar order. Direct sensitivity to the proposed order parameter is now essential. Resonant x-ray scattering, elastic-constant or ultrasound measurements, Raman scattering, and single-crystal diffuse neutron scattering would test for quadrupolar order, magnetoelastic coupling, and finite inter-bilayer correlations. DyO$X$ provides a particularly useful setting for these studies because the local magnetic building block is preserved while the coupling between bilayers is tuned chemically over a wide range.



\begin{acknowledgements}
We acknowledge Wei Zhou for assistance with the DCS experiment and Joe Paddison for assistance with the reverse Monte Carlo refinements. This project was funded by the U.S. Department of Energy, Office of Basic Energy Sciences, Materials Sciences and Engineering Division under Award DE-SC-0018660. A portion of this research used resources at the High Flux Isotope Reactor and the Spallation Neutron Source, a DOE Office of Science User Facility operated by the Oak Ridge National Laboratory. The beam time was allocated to HB-2A on proposal numbers IPTS-22419.1, IPTS-22419.2 and IPTS-27847, HYSPEC on proposal number IPTS-22418 and SEQUOIA on proposal number IPTS-22410. The x-ray diffraction measurements were performed in part at the Georgia Tech Institute for Matter and Systems, a member of the National Nanotechnology Coordinated Infrastructure, which is supported by the National Science Foundation under Grant No. ECCS-2025462.
\end{acknowledgements}



%

\clearpage 
\appendix 

\section*{\label{sec:appendices} Appendix}

\begin{table*}
\caption{\label{tab:cffit} Eigenvectors and Eigenvalues of DyOCl from the charge renormalization crystal field Hamiltonian fit in the $J$ basis. The optimal charge renormalization was found to be a factor of 1.196.}
\begin{ruledtabular}
\begin{tabular}{c|cccccccccccccccc}
E (meV) &$| -\frac{15}{2}\rangle$ & $| -\frac{13}{2}\rangle$ & $| -\frac{11}{2}\rangle$ & $| -\frac{9}{2}\rangle$ & $| -\frac{7}{2}\rangle$ & $| -\frac{5}{2}\rangle$ & $| -\frac{3}{2}\rangle$ & $| -\frac{1}{2}\rangle$ & $| \frac{1}{2}\rangle$ & $| \frac{3}{2}\rangle$ & $| \frac{5}{2}\rangle$ & $| \frac{7}{2}\rangle$ & $| \frac{9}{2}\rangle$ & $| \frac{11}{2}\rangle$ & $| \frac{13}{2}\rangle$ & $| \frac{15}{2}\rangle$ \tabularnewline
 \hline 
0.00 & 0.0 & 0.015 & 0.0 & 0.0 & 0.0 & 0.651 & 0.0 & 0.0 & 0.0 & 0.753 & 0.0 & 0.0 & 0.0 & 0.096 & 0.0 & 0.0 \tabularnewline
0.000 & 0.0 & 0.0 & -0.091 & 0.0 & 0.0 & 0.0 & -0.753 & 0.0 & 0.0 & 0.0 & -0.651 & 0.0 & 0.0 & 0.0 & -0.015 & 0.0 \tabularnewline
0.726 & -0.003 & 0.0 & 0.0 & 0.0 & 0.494 & 0.0 & 0.0 & 0.0 & 0.826 & 0.0 & 0.0 & 0.0 & 0.272 & 0.0 & 0.0 & 0.0 \tabularnewline
0.726 & 0.0 & 0.0 & 0.0 & 0.272 & 0.0 & 0.0 & 0.0 & 0.825 & 0.0 & 0.0 & 0.0 & 0.494 & 0.0 & 0.0 & 0.0 & -0.003 \tabularnewline
25.68 & 0.005 & 0.0 & 0.0 & 0.0 & -0.714 & 0.0 & 0.0 & 0.0 & 0.207 & 0.0 & 0.0 & 0.0 & 0.669 & 0.0 & 0.0 & 0.0 \tabularnewline
25.675 & 0.0 & 0.0 & 0.0 & 0.669 & 0.0 & 0.0 & 0.0 & 0.207 & 0.0 & 0.0 & 0.0 & -0.714 & 0.0 & 0.0 & 0.0 & 0.005 \tabularnewline
29.084 & 0.0 & 0.033 & 0.0 & 0.0 & 0.0 & 0.740 & 0.0 & 0.0 & 0.0 & -0.603 & 0.0 & 0.0 & 0.0 & -0.296 & 0.0 & 0.0 \tabularnewline
29.084 & 0.0 & 0.0 & -0.296 & 0.0 & 0.0 & 0.0 & -0.603 & 0.0 & 0.0 & 0.0 & 0.740 & 0.0 & 0.0 & 0.0 & 0.033 & 0.0 \tabularnewline
35.471 & -0.004 & 0.0 & 0.0 & 0.0 & 0.495 & 0.0 & 0.0 & 0.0 & -0.525 & 0.0 & 0.0 & 0.0 & 0.692 & 0.0 & 0.0 & 0.0 \tabularnewline
35.471 & 0.0 & 0.0 & 0.0 & 0.692 & 0.0 & 0.0 & 0.0 & -0.525 & 0.0 & 0.0 & 0.0 & 0.495 & 0.0 & 0.0 & 0.0 & -0.004 \tabularnewline
40.715 & 0.0 & 0.012 & 0.0 & 0.0 & 0.0 & 0.164 & 0.0 & 0.0 & 0.0 & -0.264 & 0.0 & 0.0 & 0.0 & 0.950 & 0.0 & 0.0 \tabularnewline
40.715 & 0.0 & 0.0 & 0.950 & 0.0 & 0.0 & 0.0 & -0.264 & 0.0 & 0.0 & 0.0 & 0.164 & 0.0 & 0.0 & 0.0 & 0.012 & 0.0 \tabularnewline
60.961 & 0.0 & 0.999 & 0.0 & 0.0 & 0.0 & -0.037 & 0.0 & 0.0 & 0.0 & 0.012 & 0.0 & 0.0 & 0.0 & -0.003 & 0.0 & 0.0 \tabularnewline
60.961 & 0.0 & 0.0 & -0.003 & 0.0 & 0.0 & 0.0 & 0.012 & 0.0 & 0.0 & 0.0 & -0.037 & 0.0 & 0.0 & 0.0 & 0.999 & 0.0 \tabularnewline
105.852 & 0.0 & 0.0 & 0.0 & 0.000 & 0.0 & 0.0 & 0.0 & -0.001 & 0.0 & 0.0 & 0.0 & 0.007 & 0.0 & 0.0 & 0.0 & 1.0 \tabularnewline
105.852 & -1.0 & 0.0 & 0.0 & 0.0 & -0.007 & 0.0 & 0.0 & 0.0 & 0.001 & 0.0 & 0.0 & 0.0 & -0.000 & 0.0 & 0.0 & 0.0 \tabularnewline
\end{tabular}\end{ruledtabular}
\end{table*}

\begin{figure}[h!]
    \centering
    \includegraphics[width=.35\textwidth]{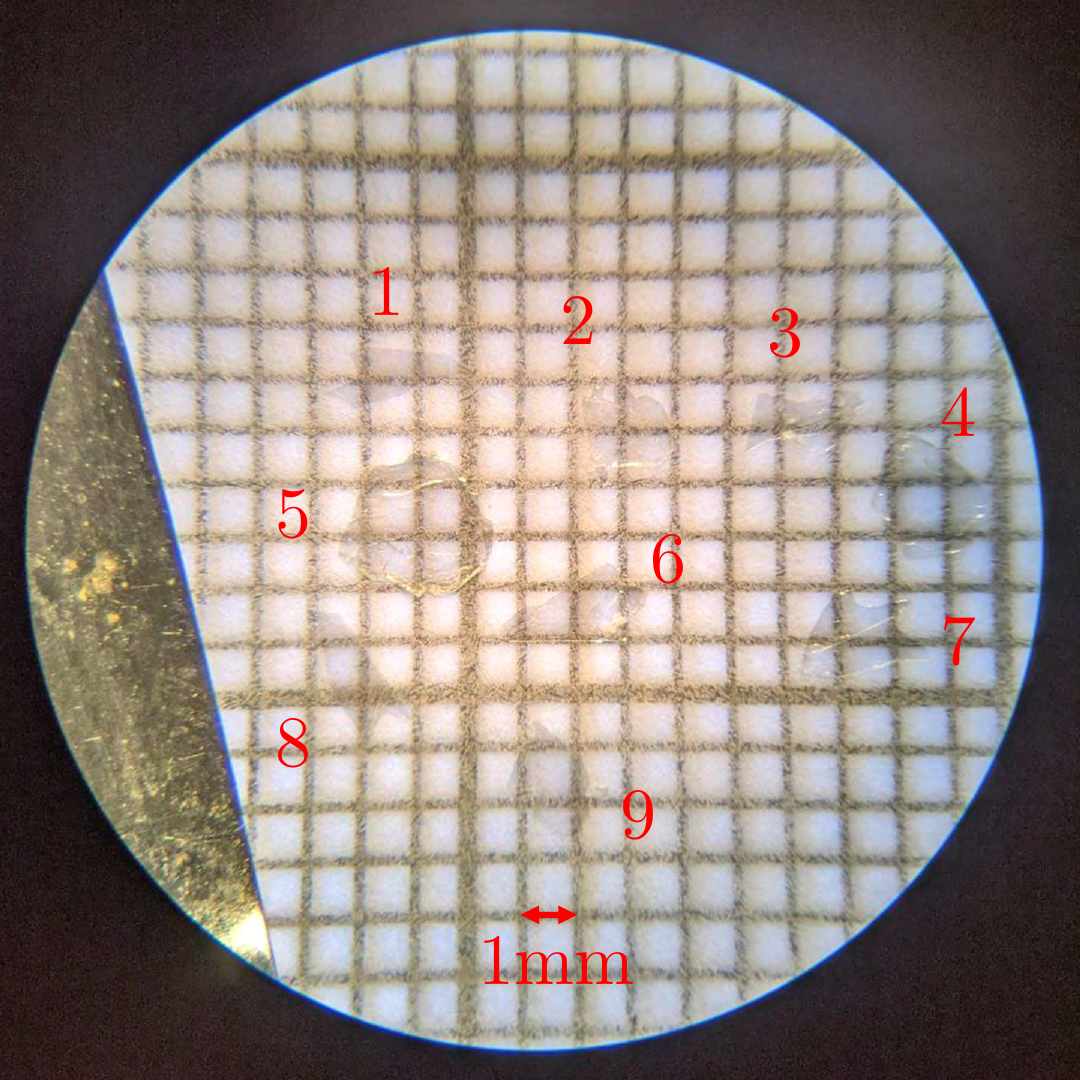}
    \caption{Single crystal samples of DyOCl.}
    \label{fig:app:crystal_pics}
\end{figure}
\begin{figure}[h!]
    \centering
    \includegraphics[width=.79\columnwidth]{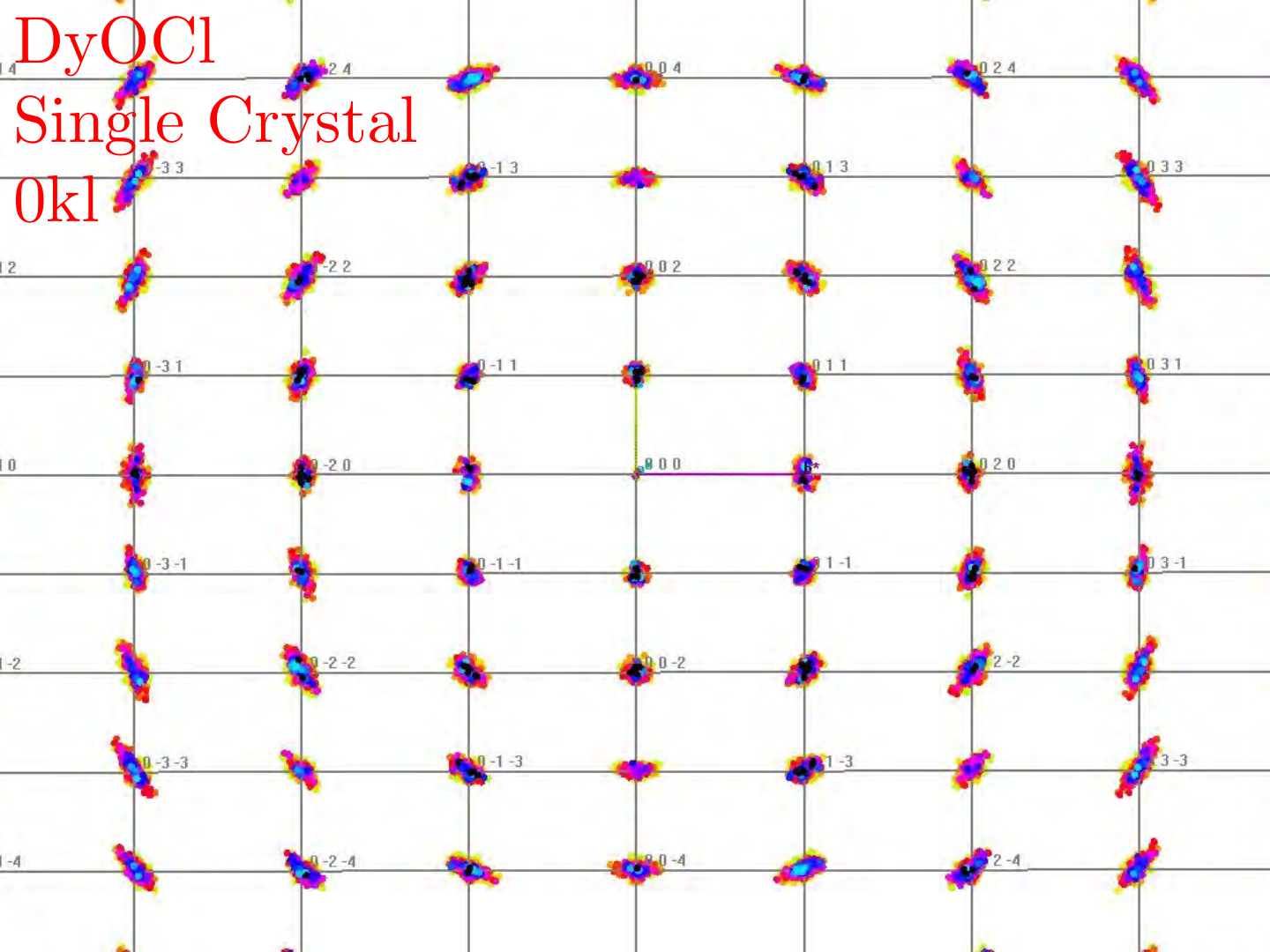}
    \includegraphics[width=.79\columnwidth]{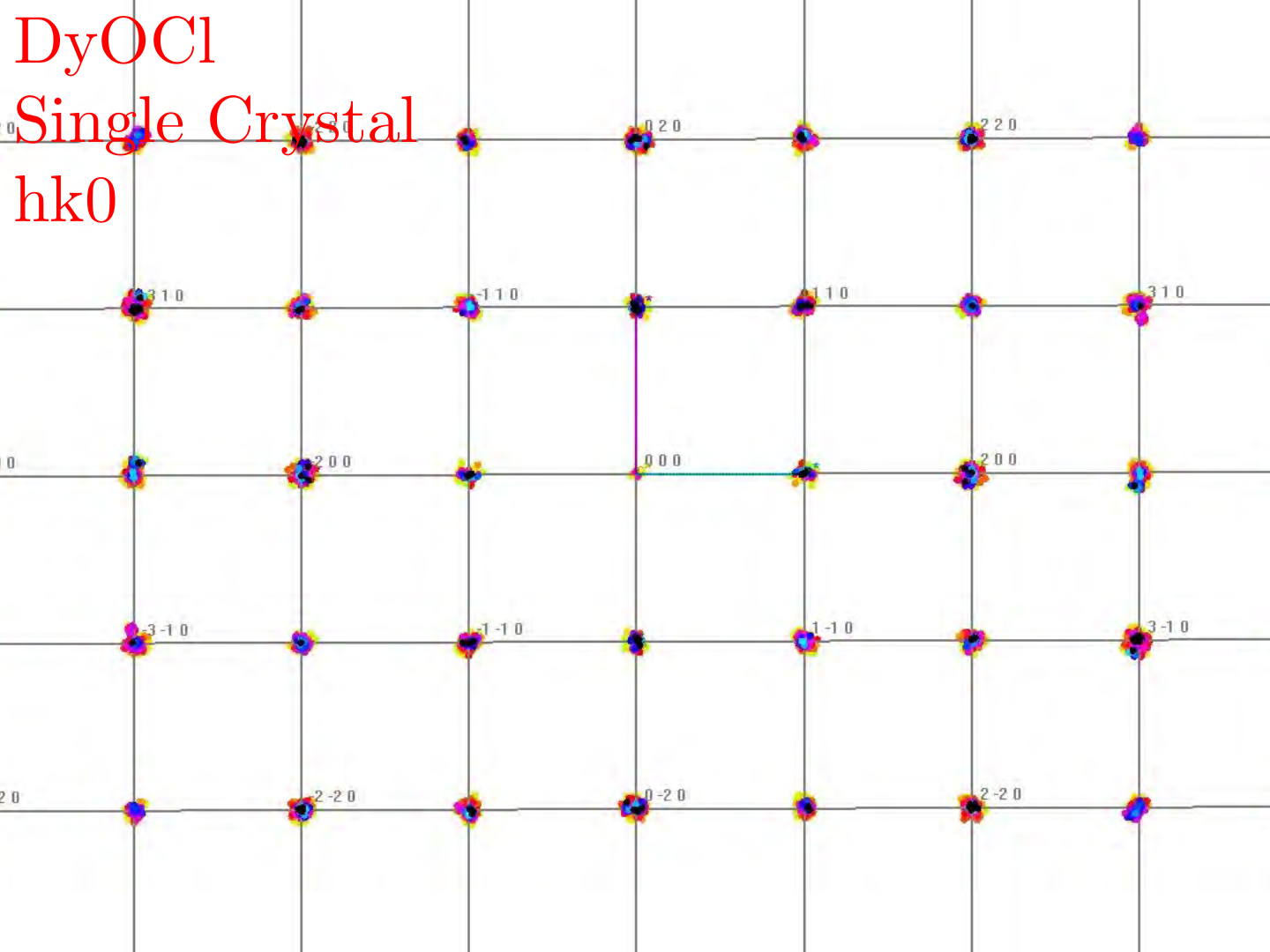}
    \caption{Reconstructed reciprocal-space planes from single-crystal x-ray diffraction measurements of DyOCl in the $(0,k,\ell)$ and $(h,k,0)$ planes.}
    \label{fig:app:crystal_scxrd}
\end{figure}

\begin{figure}[h!]
    \centering
    \includegraphics[width=.49\textwidth]{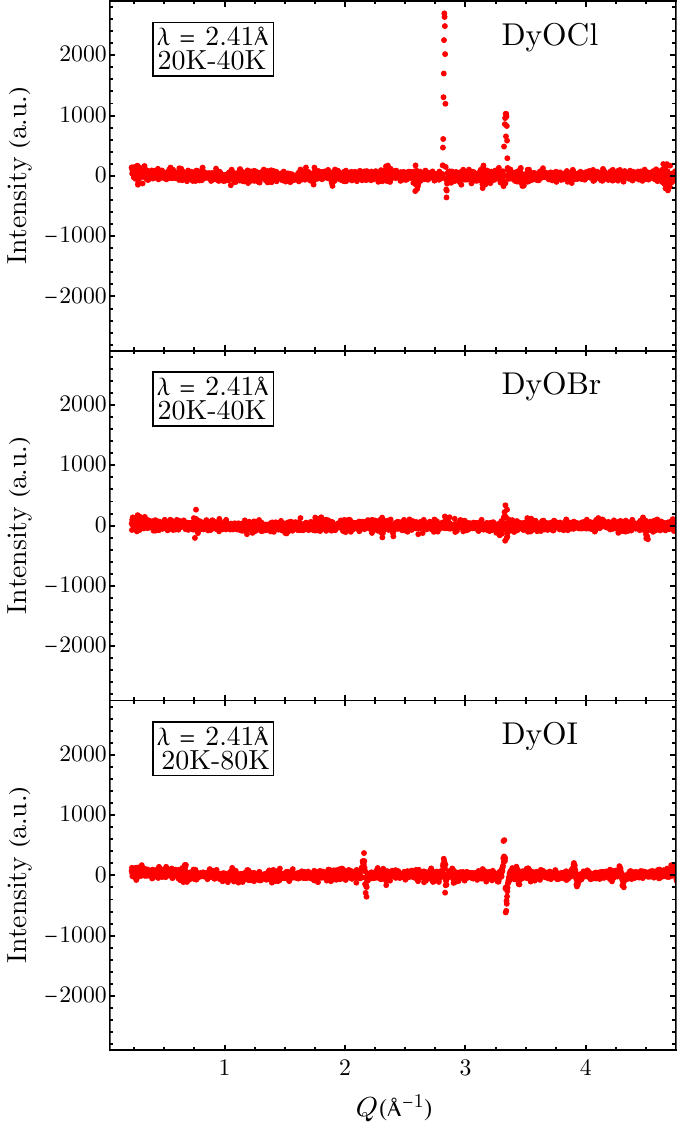}
    \caption{Neutron powder diffraction difference patterns across the higher-temperature anomaly for (a) DyOCl, (b) DyOBr, and (c) DyOI. The patterns were obtained by subtracting data collected in the paramagnetic regime at $T_{\rm pm}=40$~K for DyOCl and DyOBr or $T_{\rm pm}=80$~K for DyOI from data collected at $T=20$~K, for which $T_N<T<T_Q$. No additional magnetic Bragg reflections are resolved at $T=20$~K. The narrow positive-negative features coincide with nuclear or aluminum reflections and result predominantly from small temperature-dependent shifts of the corresponding peak positions.}
    \label{fig:app:NPDdiff}
\end{figure}
\clearpage

\end{document}